\documentclass[aps,prd,onecolumn,groupedaddress,preprintnumbers,superscriptaddress,showpacs]{revtex4}
\usepackage{amsmath,amssymb,latexsym,graphicx,multirow}

\begin{document}

\title{High-energy effects on the spectrum of inflationary gravitational
wave background in braneworld cosmology}

\author{Takashi Hiramatsu}\email{hiramatsu_at_utap.phys.s.u-tokyo.ac.jp}
\affiliation{Department of Physics, School of Science, University of Tokyo, 7-3-1 Hongo, Bunkyo, Tokyo 113-0033, Japan}

\preprint{UTAP-548} 
\preprint{RESCEU-4/06}

\date{\today}

\begin{abstract}
 We discuss the cosmological evolution of the inflationary gravitational
 wave background (IGWB) in the Randall-Sundrum single-brane model. In
 braneworld cosmology, in which three-dimensional space-like
 hypersurface that we live in is embedded in five-dimensional anti de
 Sitter (AdS$_5$) spacetime, the evolution of gravitational wave (GW)
 modes is affected by the non-standard expansion of the universe and the
 excitation of the Kaluza-Klein modes (KK-modes). These are significant
 in the high-energy regime of the universe.  We numerically evaluate
 these two effects by solving the evolution equation for GWs propagating
 through the AdS$_5$ spacetime. Using a plausible initial condition
 from inflation, we find that the excitation of KK-modes can be
 characterized by a simple scaling relation above the critical frequency
 $f_{\rm crit}$ determined from the length scale of the fifth dimension
 $\ell$. The remarkable point is that this relation generally holds as
 long as the matter content of the universe is described by the perfect
 fluid with the equation of state (EOS) $p=w\rho$ for $0\leq w\leq 1$. 
 The resultant scaling relation is
 translated into the energy spectrum of the IGWB as 
 $\Omega_{\rm GW}\propto f^{(3w-1)/(3w+2)}$ for $f>f_{\rm crit}$. This
 indicates that in the radiation dominant case ($w=1/3$), the two
 high-energy effects accidentally compensate each other and the spectrum
 becomes almost the same as the one predicted in the
 four-dimensional theory, i.e., $\Omega_{\rm GW}\propto f^0$.
\end{abstract}

\pacs{04.30.-w, 04.30.Nk, 04.50.+h, 98.80.-k}

\maketitle

\section{Introduction}
\label{sec:introduction}

Gravitational waves (GWs) are ultimate probes of the untouched region
of the universe. Currently, large scale ground-based interferometers
(TAMA300\cite{Ando2005}, LIGO\cite{Sigg2004}, VIRGO\cite{Acernese2005},
GEO600\cite{Grote2005}) are enthusiastically trying to detect the 
signals emitted from stellar objects with relativistic motion --
supernovae explosions, coalescence of neutron star binaries and so
on. Among numerous types of GWs, the gravitational wave background (GWB) may
possess much interesting information on the cosmology, though its
detection is expected to be so challenging
\cite{Abbott2005GWB}. Especially, the inflationary GWB (IGWB), generated
during the inflationary epoch by the quantum fluctuations of the
spacetime\cite{Starobinsky1979, Rubakov1982, Fabbri1983, Abbott1984a,
Abbott1984b, Abbott1986, Allen1988, Turner1997, Maggiore}, is
thought to be one of the most fundamental predictions of the
inflationary scenario \cite{Sato1981, Guth1981, Linde1982, Albrecht1982}. 
Since the history of the cosmological expansion is imprinted in the
power spectrum of the IGWB, it helps us to understand the extremely early
universe if we can detect the signals by the future space-based
experiments, such as DECIGO\cite{DECIGO} and BBO\cite{BBO, Ungarelli2005}.

As an ultimate cosmological tool, the IGWB may also be useful to probe
the presence of extra-dimensional spaces. Recent developments in
particle physics suggest that we live in a higher dimensional
spacetime. In particular, braneworld scenarios have recently attracted
much attention theoretically and observationally (for a review, see
\cite{rev_Maartens}). According to them, we live in a three-dimensional
hypersurface (brane) embedded in the higher-dimensional spacetime (bulk).
While gravity can propagate in the bulk with the curvature scale $\ell$,
the standard model particles are confined to the three-dimensional
brane. In the low-energy regime of the universe ($H\ell \ll 1$),
four-dimensional general relativity is successfully recovered and the
extra-dimensional effects should be fairly small. On the other hand, in
the high-energy regime ($H\ell \gg 1$), the localization of gravity is
not always guaranteed and a significant deviation of the time evolution
of GWs from the standard four-dimensional theory is expected.
If this scenario is true, the spectrum of the IGWB may be significantly
modified by the high-energy effects, which can provide a direct probe of
the extra-dimensions.

The goal of this paper is to investigate the high-energy effects on
the evolution of the IGWB and quantify the power spectrum. During the
inflationary epoch, the wavelength of the GWs exceeds the Hubble horizon
scale due to the exponential expansion of the universe and the amplitude
of GWs becomes frozen. After the end of inflation, the universe enters
the decelerated expansion phase and wavelengths of GWs soon become
shorter than the Hubble scale. When the Hubble horizon
scale becomes comparable or smaller than the characteristic size of the
extra-dimension, the high-energy effects may significantly
affect the evolution of GWs. There are two main high-energy effects :
i) peculiar cosmological expansion due to the high-energy correction
of the Friedmann equation, which enhances the spectrum in the high
frequency region and ii) excitation of Kaluza-Klein modes (KK-modes) freely
escaping from the brane to the bulk spacetime, which may suppress the
amplitude of the GWs on the brane. While the former effect is simply
estimated from the expansion rate of the universe, the amount of the
latter effect requires the knowledge of the wave propagation in the
bulk. Focusing on the Randall-Sundrum single-brane model, in which the
three-dimensional hypersurface (brane) is embedded in the
five-dimensional anti-de Sitter (AdS$_5$) spacetime\cite{RS2}, 
many authors have tried to estimate these effects in an analytic way.
However, these analyses were restricted to the idealistic situations
\cite{Gorbunov,Kobayashi,Kobayashi2004} or low-energy cases in the
Friedmann universe \cite{Battye,Easther}, since the analytic study of
wave equation is generally intractable due to the complicated form of
equation as well as boundary condition. Thus, in this paper, we
numerically solve the wave equation and try to estimate the observed
IGWB spectrum.

Concerning numerical studies, several authors have used different
numerical techniques and coordinate systems to solve the wave equation
of GWs
\cite{Hiramatsu1,Hiramatsu2,Ichiki,Kobayashi2005A,Kobayashi2005B,Sanjeev2006}.
In our previous studies \cite{Hiramatsu1, Hiramatsu2}, numerical
simulations were carried out in the two types of coordinate systems. 
One is the Gaussian normal coordinate system in which we found
the suppression of the amplitude of the IGWB on the brane. Unfortunately, 
the coordinate singularity appears in the bulk and this restricts our
analyses to a relatively low-energy scale \cite{Hiramatsu1}. On the
other hand, another coordinate system we used is the Poincar\'e
coordinates in which we observed that the two high-energy effects
compensate each other and the spectrum became same one as predicted in
the four-dimensional theory \cite{Hiramatsu2}.

In this paper, we first show that the spectrum of the IGWB estimated in
the paper \cite{Hiramatsu2} is robust against several numerical
artifacts, such as the dependence of the regulator brane and initial
time. Next, we study the dependence of the spectra on the equation of
state (EOS) of the universe to clarify how the two high-energy effects
change with the expansion rate of the universe.

This paper consists of seven sections. In Sec. 2, we discuss how the
spectrum of the IGWB is affected by the presence of the
extra-dimensions. In Sec. 3, we introduce the Poincar\'e coordinate
system and derive the evolution equation of GWs. After briefly
discussing the details of the numerical simulations  and initial
conditions in Sec. 4, we check our numerical scheme by solving a simple
case in Sec. 5. In Sec. 6, we present the numerical results in the case
that the IGWB re-enters the Hubble horizon during the radiation dominated
(RD) universe. Also in Sec. 6, the results in cases with other equation
of state (EOS) are
shown. Finally, Sec. 7 is devoted to the summary and conclusions.

In this paper, we use units in which $c=\hbar=1$.
$M_\text{pl}$ represents the four-dimensional Planck mass/energy.

\section{Gravitational Wave Background from Inflation}
\label{sec:GWB}

%
\subsection{Standard four-dimensional prediction}
\label{subsec:4D}
%

Standard inflation model predicts that gravitational waves (GWs) are
generated by the quantum fluctuations of spacetime. In this section,
we briefly review how to evaluate the power spectrum of IGWB
(see also Ref.\cite{Maggiore}).

\begin{figure}[ht]
 \centering
 {
   \includegraphics[width=8cm]{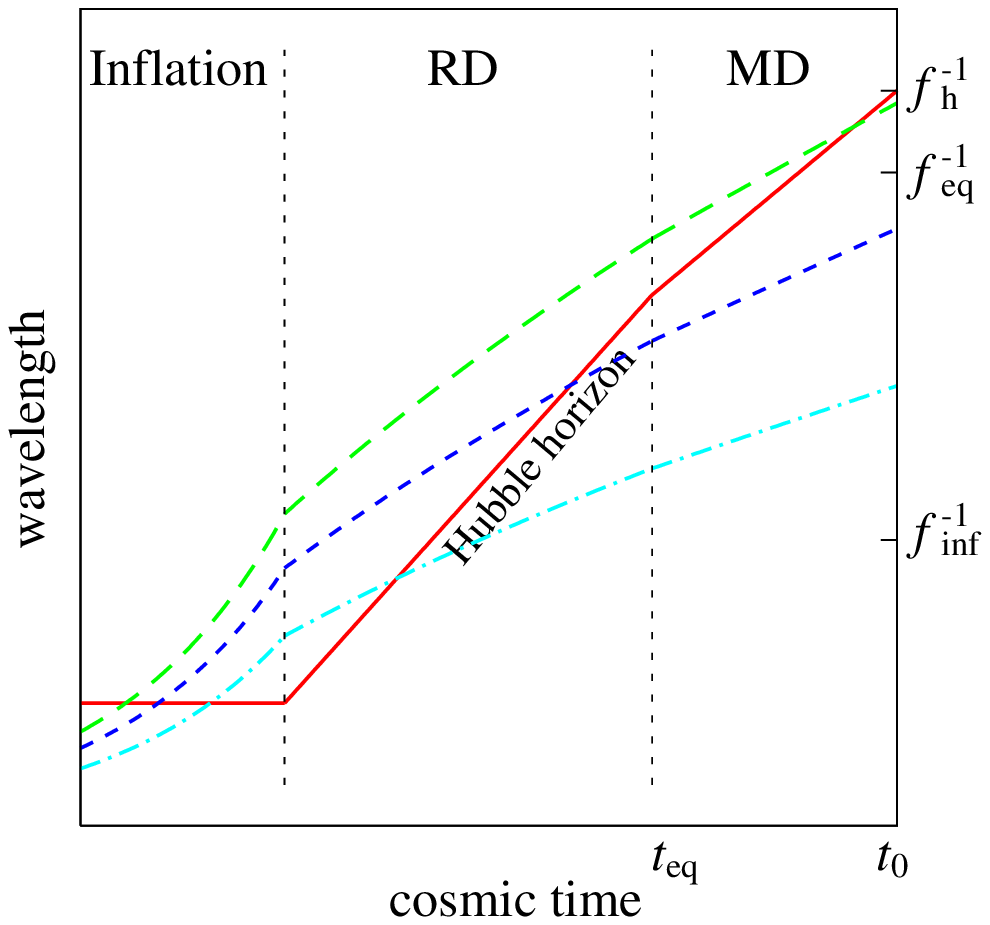}
   \includegraphics[width=8cm]{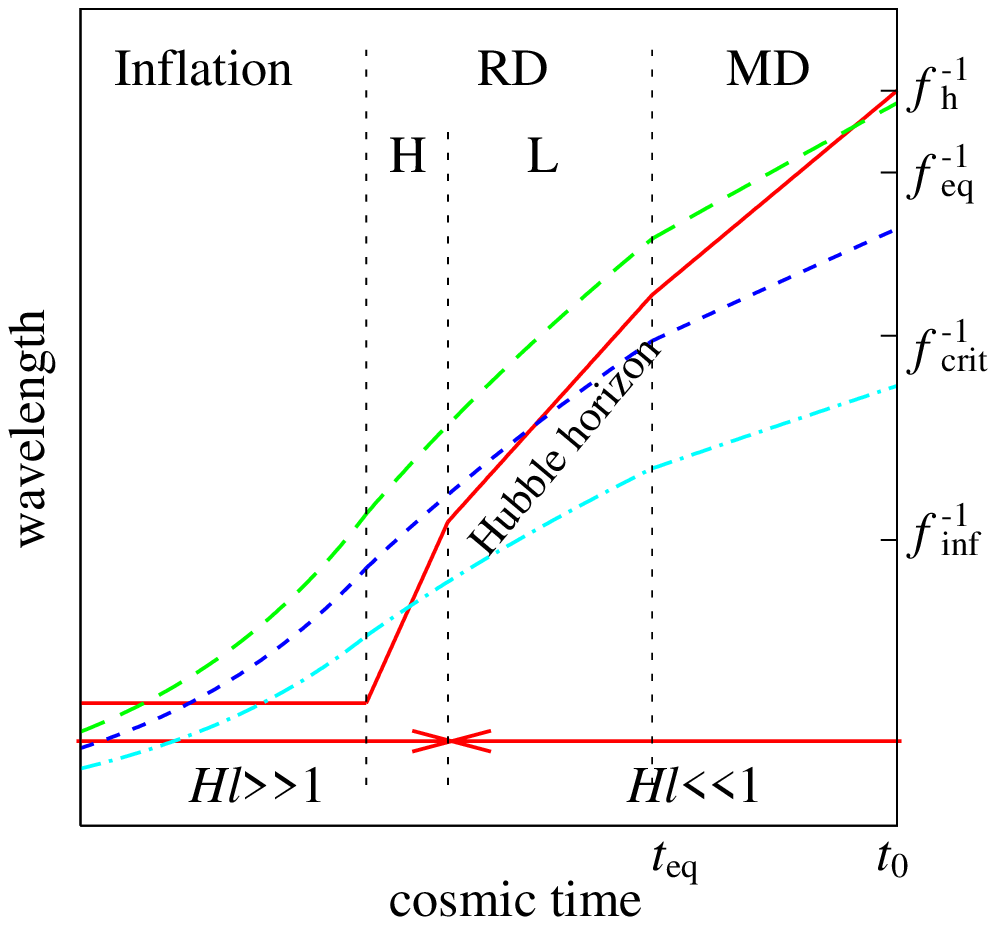}
 }
 \caption{Schematic diagrams of the evolution histories of GWs.
 {\it Left} : four-dimensional case, {\it Right} : five-dimensional
 case. In the high-energy RD regime of the latter case (denoted by ``H''),
 the $\rho^2$-term in the Friedmann equation changes the time dependence of
 the Hubble parameter. This fact yields a new characteristic frequency
 $f_{\rm crit}$.
 \label{fig:schematic_diagram}} 
\end{figure}

In the left panel of Fig.\ref{fig:schematic_diagram}, a sketch of the 
evolution histories of GWs with various wavelengths is shown. The
vertical and the horizontal axes represent the wavelength of GWs and the
cosmic time, respectively. In this figure, the solid line represents the
Hubble horizon scale $H^{-1}$, and we have labeled three regimes 
as ``Inflation'', ``RD'' and ``MD'' for the
inflationary epoch, the radiation-dominated epoch and the
matter-dominated epoch, respectively. During the inflation, the universe
experiences accelerated expansion and the wavelength of GWs eventually exceeds
the Hubble horizon scale.  Then the oscillatory behavior ceases to exist
and the amplitudes of GWs become frozen. After inflation, these GWs
re-enter the horizon in the decelerated expansion phase
(regions ``RD'' and ``MD''). Inside the horizon, the wavelengths are
redshifted and the amplitudes are reduced by the cosmological expansion.
Since the horizon re-entry time depends on the comoving wave number for
each GW mode, the resultant energy spectrum of the IGWB observed at
present simply reflects the expansion rate at the horizon
re-entry time.

To evaluate the spectrum of IGWB, we first consider the characteristic
frequencies of GWB associated with the cosmic history. We define three
characteristic frequencies according to the standard four-dimensional
cosmology : i) the lowest frequency $f_{\rm h}$; ii) the frequency of GWs
re-entering the horizon just at the matter-radiation equality time,
$f_{\rm eq}$; and iii) the cut-off frequency by the inflation, $f_{\rm inf}$.
First, the largest wavelength of IGWB observed today is definitely the
horizon length which corresponds to the frequency
%
\begin{equation}
f_{\rm h}\approx 2.3\times 10^{-18}\;{\rm Hz}
\left(\frac{H_0}{72\;{\rm km}/{\rm s}\cdot{\rm Mpc}}\right),
\end{equation}
%
where $H_0$ denotes the present value of the Hubble parameter.
Second,
the frequency of large-scale GWs which came into the Hubble
horizon at the matter-radiation equality time $t_{\rm eq}$ can be
calculated as
%
\begin{equation}
 f_{\rm eq} = \frac{1}{2\pi}\frac{a_{\rm eq}}{a_{0}}H_{\rm eq}
    \approx 2.1\times 10^{-17}\;{\rm Hz}\;
\left(\frac{H_0}{72\;{\rm km}/{\rm s}\cdot{\rm Mpc}}\right)
\left(\frac{1+z_{\rm eq}}{3200}\right)^{1/2}, \label{eq:f_eq_4D}
\end{equation}
%
where $a$ denotes the scale factor and $z$ the redshift. The subscripts
'0' and 'eq' represent the quantities evaluated at the present time
$t_0$ and at the matter-radiation equality time $t_{\rm eq}$,
respectively. Finally, the highest frequency observed today is
determined from the Hubble horizon at the end of the inflation, which
can be calculated in the same way as (\ref{eq:f_eq_4D}): 
%
\begin{equation}
 f_{\rm inf} \approx \frac{1}{2\pi}\frac{a_{\rm inf}}{a_{\rm eq}}
               \frac{a_{\rm eq}}{a_{0}}H_{\rm inf}
    = 1.1\;{\rm GHz} 
      \left(\frac{H_{\rm inf}}{6\times 10^{-5}M_{\rm pl}}\right)^{1/2}
      \left(\frac{H_0}{72\;{\rm km}/{\rm s}\cdot{\rm Mpc}}\right)^{1/2}
      \left(\frac{1+z_{\rm eq}}{3200}\right)^{-1/4},
 \label{eq:f_inf_4D}
\end{equation}
%
where $H_{\rm inf}$ means the energy scale of the inflation, which is
constrained by the COBE observation as 
$H_{\rm inf} < 6\times 10^{-5}M_{\rm pl}$ \cite{Maggiore}.
As a consequence, the GWs with $f_{\rm h}<f<f_{\rm eq}$ re-enter the
Hubble horizon during the MD phase, while for
$f_{\rm eq}<f<f_{\rm inf}$, GWs re-enter during the RD phase. These
characteristic frequencies are shown in Fig. \ref{fig:schematic_diagram}. 

Let us focus on the shape of the IGWB spectrum.
Conventionally, the power spectrum of GWB is characterized by the energy
density instead of its amplitude. We introduce the quantity
$\Omega_{\rm GW}$ defined as \cite{Maggiore}
%
\begin{equation}
 \Omega_{\rm GW}(f) \equiv 
  \frac{1}{\rho_{\rm c}}\frac{d\rho_{\rm GW}}{d\log f},
\end{equation}
%
where $\rho_{\rm c}=3H_0^2/8\pi G=9.8\times 10^{-30}$ g$/$cm$^3$
means the critical density of the universe and $\rho_{\rm GW}$ the
energy density of GWB. Denoting the characteristic
amplitude by $h$, the above quantity is related to
%
\begin{equation}
 h_0^2\Omega_{\rm GW} = \left(\frac{h}{1.263\times
		 10^{-18}}\right)^2\left(\frac{f}{1{\rm Hz}}\right)^2.
 \label{eq:Omega_propto_h}
\end{equation}
%

The frequency dependence of the power spectrum can be derived from the
fact that the amplitude of the GWs evolves as $h\propto 1/a$ inside the
horizon. Assuming the scale factor evolves as $a\propto t^n$ near the
horizon re-entry time $t_*$, the GW amplitude observed today is related
to $t_*$ as 
%
\begin{equation}
 h_{0} = \frac{a_*}{a_{0}}h_* \propto t_*^nf^{n_{\rm T}/2}.
 \label{eq:h_propto_t}
\end{equation}
%
Here $h_*$ denotes the amplitude of GWs evaluated at the time $t_*$.
The amplitude $h_*$ is primarily determined by the quantum fluctuations
generated during the inflation, whose spectral dependence is given by
$h_*^2 \propto f^{n_{\rm T}}$ (e.g., Sec. 6.5 of Ref. \cite{LiddleLyth}).
In a single-field model of slow-roll inflation, the spectral index
$n_{\rm T}$ can be expressed by the slow-roll parameter $\epsilon$ as
$n_{\rm T}\approx-2\epsilon$ \cite{Smith2005}\cite{LiddleLyth}. In the pure
de Sitter expansion, $n_{\rm T}=0$.

In the power-law expansion, the Hubble parameter at $t=t_*$ scales as
%
\begin{equation}
  H_*\propto t_*^{-1}. 
  \label{eq:H_propto_t}
\end{equation}
%
Hence the observed frequency $f$ is related to $t_*$ as
%
\begin{equation}
 f = \frac{k}{2\pi a_{0}}= \frac{a_*H_*}{2\pi a_{0}} 
      \propto t_*^{n-1},
 \label{eq:f_propto_t}
\end{equation}
%
where $k=a_*H_*$ denotes the comoving wave number of the GW concerned.
From (\ref{eq:h_propto_t}) and (\ref{eq:f_propto_t}), the power
spectrum of the IGWB becomes
%
\begin{equation}
 \Omega_{\rm GW} \propto h_0^2f^2 \propto f^{\frac{4n-2}{n-1}+n_{\rm T}}.
 \label{eq:O_propto_f}
\end{equation}
%
Particularly in cases with the matter content in the universe satisfying
the equation of state (EOS),
%
\begin{equation}
p = w\rho,
\label{eq:EOS}
\end{equation}
%
the power-law index of the scale factor, $n$, is rewritten with
%
\begin{equation}
 n = \frac{2}{3(1+w)}.
 \label{eq:scalefactor_index_4D}
\end{equation}
%
Then the energy spectrum (\ref{eq:O_propto_f}) becomes
%
\begin{equation}
 \Omega_{\rm GW} \propto \displaystyle f^{\frac{6w-2}{3w+1}+n_{\rm T}}.
 \label{eq:Omega_w_4D}
\end{equation}
%
Now, simply assuming $n_{\rm T}=0$ and applying this formula to the 
standard history of the universe [RD ($w=1/3$) phase and MD ($w=0$) phase],
the energy spectrum of the IGWB observed at present becomes \cite{Turner1997}
%
\begin{equation}
 \Omega_{\rm GW} \propto
 \begin{cases}
  f^{0} & (f_{\rm eq} < f < f_{\rm inf}), \\
  f^{-2} &(f_{\rm h} < f < f_{\rm eq}).
 \end{cases}
 \label{eq:O_propto_f_specific}
\end{equation}
%
The resultant spectrum in the four-dimensional cosmology is shown in
long-dashed lines in Fig. \ref{fig:schematic_spectrum}. The
normalization of the spectrum is weakly constrained from the CMB
observation by COBE, which leads to $\Omega_{\rm GW} < 10^{-14}$
\cite{Maggiore}. This figure shows that the IGWB widely exists over the
frequencies which spans approximately 30 orders of
magnitude. Furthermore most of the frequency region comes from the GWs
which re-enters the horizon during the RD phase. 

%
\subsection{High-energy effects in the braneworld cosmology}
\label{subsec:RS}
%

In a braneworld scenario, the propagation of gravity may be modified by
the presence of extra-dimensional spaces, which affects the observed
spectrum of IGWB. In this paper, we consider the Randall-Sundrum
single-brane model \cite{RS2}. In this model, a three-dimensional brane
is embedded in five dimensional anti-de Sitter spacetime (AdS${}_5$
bulk) with the curvature scale $\ell$.  Here we consider the flat
Friedmann-Robertson-Walker (FRW) universe on the brane. In general, the
RS model may possess a black-hole in the bulk, but we do not consider it.

%
\subsubsection{Cosmological Expansion on the Brane}
%

In the RS model, the $\mathbb{Z}_2$ symmetry (mirror
symmetry) is imposed on the brane, whose physical meaning comes from the
$S^1/\mathbb{Z}_2$ orbifold compactification in heterotic M-theory. This
symmetry provides the junction condition which gives a relation
between the extrinsic curvature $K_{\mu\nu}$ at the brane and the
energy density of matter on the brane. From the junction condition, the
Friedmann equation on the brane and the conservation law become
\cite{Mukohyama1999a,Binetruy1999a,Binetruy1999b,Langlois2000,Kraus,Ida}:
%
\begin{equation}
 H^2(t) = \left(\frac{\dot{a}}{a}\right)^2 = \frac{\kappa_4^2}{3}
\rho\left(1+\frac{\rho}{2\lambda}\right),
\qquad 
 \dot{\rho} = -3(\rho+p)H,
\label{eq:Friedmann}
\end{equation}
%
where $\kappa_4^2$ represents the gravity constant defined as
$\kappa_4^2 = 8\pi G = 8\pi/M_\text{pl}^2$ and $\lambda$ denotes the
tension of the brane. The tension is related with the bulk curvature scale
$\ell$ as $\kappa_4^2\lambda\ell^2=6$ if we require that the cosmological
constant on the brane $\Lambda_4$ vanishes. For later convenience, 
we define a dimensionless energy density normalized by the tension
$\epsilon(t)$ by $\epsilon(t)\equiv\rho(t)/\lambda$.

From (\ref{eq:Friedmann}), we can separate the regime into two parts :
the high-energy regime ($H\ell>1$) and the low-energy regime ($H\ell<1$).
The critical energy density satisfying the relation $H\ell=1$ is given by
%
\begin{equation}
\epsilon_{\rm crit} = \sqrt{2}-1.
\label{eq:critical_energy}
\end{equation}
%
Thus, when $\epsilon(t) \gg \epsilon_{\rm crit}$, the high-energy
correction characterized by the $\rho^2$-term in (\ref{eq:Friedmann})
significantly modifies the cosmological expansion.

In the cosmology with perfect fluid, the exact solutions for the scale
factor $a(t)$ and the normalized energy density $\epsilon(t)$ are known.
These solutions are expressed as (e.g., \cite{Binetruy1999a, Binetruy1999b}):
%
\begin{equation}
        a(t) = \left(\frac{\epsilon_{\rm crit}}{\epsilon}\right)^{1/3(1+w)},
\quad 
  \epsilon(t)=\frac{2}{\left\{3(1+w)t/\ell+1\right\}^2-1},
\label{eq:scalefactor_energy}
\end{equation}
%
where the scale factor is normalized to unity at the critical 
energy, $\epsilon_{\rm crit}$. In cases with $w=0$ (MD),
$w=1/3$ (RD) and $w=1$, the energy densities become respectively 
%
\begin{equation}
 \epsilon(t) =
 \begin{cases}
  \displaystyle \frac{2\ell^2}{9t^2+6t\ell} &\;{\rm for}\;\; w=0,\\
  \displaystyle \frac{\ell^2}{8t^2+4t\ell}  &\;{\rm for}\;\; w=1/3,\\
  \displaystyle \frac{\ell^2}{18t^2+6t\ell} &\;{\rm for}\;\; w=1.
  \end{cases}
\label{eq:energy}
\end{equation}
%
Using the above relations, we will consider how the high-energy regime
of the universe can modify the spectrum of the IGWB.

%
\subsubsection{High-energy effects on GWs}
\label{subsubsec:high-energy}
%

\begin{figure}[ht]
 \centering
 \includegraphics[width=8cm]{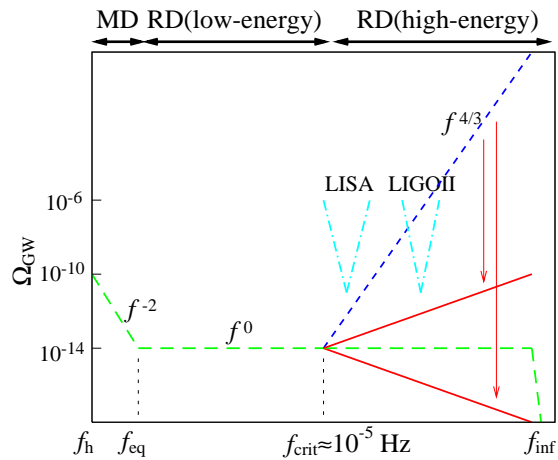}
 \caption{A schematic diagram of the spectrum of IGWB. The
 four-dimensional prediction is shown in long-dashed lines.
 Considering the modification due to the non-standard cosmological
 expansion, the spectrum behaves as $\Omega_{\rm GW}\propto f^{4/3}$
 shown in short-dashed line, which may appear upon the detection limit
 of advanced LIGO (dot-dashed line). Moreover, KK-mode excitations
 modify the spectrum as solid lines. The main issue in this paper
 is to clarify whether the resultant spectrum (solid) exceeds the
 four-dimensional prediction (long-dashed). 
 \label{fig:schematic_spectrum}} 
\end{figure}

As we mentioned in Sec. \ref{sec:introduction}, there are two
important effects on the spectrum in the high-energy regime. 
Let us first consider the non-standard cosmological expansion in presence of
$\rho^2$-term. From (\ref{eq:scalefactor_energy}), the power-law index
of the scale factor in the high-energy regime (`H' in the right panel of
Fig. \ref{fig:schematic_diagram}) is related to the EOS parameter $w$ as
%
\begin{equation}
 n= \displaystyle \frac{1}{3(1+w)} \;\;{\rm for}\;\; t\ll\ell.
 \label{eq:scalefactor_index}
\end{equation}
%
On the other hand, at the late-time phase 
($\epsilon \ll \epsilon_{\rm crit}$), the power-law index just coincides
with the four-dimensional result (\ref{eq:scalefactor_index_4D}). As a
result, the energy spectrum (\ref{eq:O_propto_f}) is modified to 
%
\begin{equation}
 \Omega_{\rm GW}=
 \begin{cases}
  \displaystyle f^{\frac{6w-2}{3w+1}} & {\rm for}\;\;f\ll f_{\rm crit},\\
  \displaystyle f^{\frac{6w+2}{3w+2}} & {\rm for}\;\;f\gg f_{\rm crit},
 \end{cases}
 \label{eq:Omega_w}
\end{equation}
%
where $f_{\rm crit}$ denotes the critical frequency given by
%
\begin{equation}
 \begin{aligned}
 f_{\rm crit} &= \frac{1}{2\pi\ell}\frac{a_{\rm crit}}{a_{\rm
  eq}}\frac{a_{\rm eq}}{a_{0}}\\
  &= 5.6\times 10^{-5}\;{\rm Hz}
  \left(\frac{\ell}{0.1\;{\rm mm}}\right)^{-1/2}
  \left(\frac{H_0}{72\;{\rm km}/{\rm s}\cdot{\rm Mpc}}\right)^{1/2}
  \left(\frac{1+z_{\rm eq}}{3200}\right)^{-1/4}.
 \end{aligned}
\label{eq:critical_frequency}
\end{equation}
%
That is, the wavelength at the horizon re-entry time just coincides with
the curvature scale, namely, $H_*=\ell^{-1}$ (cf. \cite{Hogan}). Note
that the curvature radius $\ell$ is constrained to $\ell < 0.1$ mm by
table-top experiments on the Newton force law \cite{Chiaverini,Long}. 
For the frequency $f>f_{\rm crit}$, GWs re-enter the horizon during the
RD phase of the $\rho^2$-term dominated epoch (right panel of
Fig. \ref{fig:schematic_diagram}). In the high-energy RD phase, the
spectrum of the IGWB is modified to
%
\begin{equation}
 \Omega_{\rm GW} \propto
  f^{4/3} \qquad (f_{\rm crit} < f < f_{\rm inf}),
 \label{eq:O_propto_f_specific_5D}
\end{equation}
%
which is shown in short-dashed lines in Fig
\ref{fig:schematic_spectrum}. Additionally, the inflationary cut-off
frequency (\ref{eq:f_inf_4D}) is modified to 
%
\begin{equation}
 \begin{aligned}
 f_{\rm inf} &\approx \frac{1}{2\pi}\frac{a_{\rm inf}}{a_{\rm crit}}
               \frac{a_{\rm crit}}{a_{\rm eq}}
               \frac{a_{\rm eq}}{a_{0}}H_{\rm inf}\\
    &= 4.7\times 10^6\;{\rm GHz} 
  \left(\frac{H_{\rm inf}}{6\times 10^{-5}M_{\rm pl}}\right)^{3/4}
  \left(\frac{H_0}{72\;{\rm km}/{\rm s}\cdot{\rm Mpc}}\right)^{1/2}
  \left(\frac{\ell}{0.1\;{\rm mm}}\right)^{1/4}
  \left(\frac{1+z_{\rm eq}}{3200}\right)^{-1/4}.
  \end{aligned}
\label{eq:f_inf_5D}
\end{equation}
%

Notice that the above estimate neglects another remarkable high-energy
effect caused by the excitation of Kaluza-Klein modes (KK-modes).
The KK-modes can propagate into the bulk and may be observed on the
brane as massive gravitons. By contrast, the GW propagating on the brane
is called the 'zero-mode', and it behaves like a massless graviton on the
brane. Strictly speaking, the KK-modes and zero-mode are
coordinate-dependent concepts and are mathematically well-defined
only in the case of the Minkowski brane and the de-Sitter brane.
Nevertheless, we keep to use these terms in the FRW case to distinguish
these propagation features.

As we will see in the next section, the brane is generally moving in the
bulk spacetime. From the analogy of the moving
mirror problem \cite{BirrelDavies}, even if only the zero-mode is
generated in the inflationary epoch, the zero-mode is partially
transferred to KK-modes with the arbitrary masses. If this is true, the
amplitude of the zero-mode observed on the brane may be suppressed in
comparison with the result neglecting the KK-modes, i.e.,
(\ref{eq:O_propto_f_specific_5D}). The envisaged spectrum involving the
two effects is schematically shown as solid lines in
Fig. \ref{fig:schematic_spectrum}. Unfortunately, we cannot fully
estimate the KK-mode effects in an analytic way. Thus, in order to
construct the IGWB spectrum including the KK-mode effects, we must
directly solve the evolution equations of GWs in the five-dimensional
cosmology.

In the next section, we present the formalism to solve the evolution
equations for GWs numerically.

\section{Evolution equation and initial conditions}
\label{sec:evolution_initial}

%
\subsection{Evolution equation of GWs}
\label{subsec:evolution}
%

\begin{figure}[ht]
 \centering
 {
   \includegraphics[width=6cm]{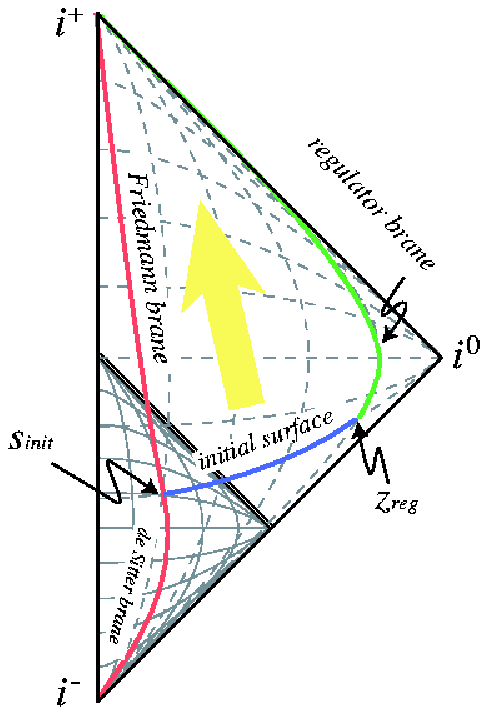}
 }
 \caption{The motion of the de Sitter and Friedmann brane in the AdS$_5$
 bulk in the Poincar\'e coordinate system.
 \label{fig:Poincare}} 
\end{figure}

Let us consider the tensor perturbations in the AdS$_5$ spacetime. 
In the Poincar\'e coordinate $(\tau,\mathbf{x},z)$, the perturbed metric 
of the AdS$_5$ spacetime is given by
%
\begin{equation}
  ds^2 = \left(\frac{\ell}{z}\right)^2
         \{-d\tau^2 + (\delta_{ij}+h_{ij})dx^idx^j+dz^2\},
  \qquad (i,j=1,2,3),
\label{eq:Poincare_perturbed_metric}
\end{equation}
where $h_{ij}$ satisfies the transverse-traceless (TT) condition,
%
\begin{equation}
 \partial_ih^i{}_j = h^i{}_i=0.
\label{eq:TTgauge}
\end{equation}
%
While this coordinate system is free from the coordinate singularities,
the brane is non-static, moving in the AdS$_5$ spacetime
\cite{Kraus,Ida}. The trajectory of the moving brane is determined from
the scale factor on the brane, which is described as
$(\tau_{\rm b},\,z_{\rm b})$:
%
\begin{equation}
 \tau_{\rm b} = T(t),\qquad z_{\rm b} = \frac{\ell}{a(t)}, 
 \label{eq:brane_trajectory} 
\end{equation}
%
where we use the cosmic time $t$ on the brane as a parameter of the
trajectory. The function $T(t)$ is given by \cite{Deffayet}
%
\begin{equation}
 \dot{T}(t) = \frac{1}{a}\sqrt{1+(H\ell)^2}. \label{eq:timefunc} 
\end{equation}
%
In Fig. \ref{fig:Poincare}, we show the trajectory denoted by
'Friedmann brane' in the conformal chart where the surfaces of $\tau=$
const. and $z=$ const. are plotted as dashed lines. One can check that
this trajectory induces the metric of the four-dimensional flat
Friedmann-Robertson-Walker model on the brane : 
$ds_\text{b} = -dt^2+a^2(t)\delta_{ij}dx^idx^j$.
Note that, in the case of de Sitter brane, the scale factor becomes 
$a(t)=e^{Ht}$ and the Hubble parameter $H$ is constant.
Hence the trajectory becomes
%
\begin{equation}
 \tau^{\rm dS}_{\rm b} = -\frac{\sqrt{1+(H\ell)^2}}{H}e^{-Ht}, \qquad 
 z^{\rm dS}_{\rm b}= \ell e^{-Ht},
 \label{eq:DS_trajectory} 
\end{equation}
%
which yields the straight trajectory in the bulk; that is, 
$\tau^{\rm dS}_{\rm b}$ becomes proportional to $z^{\rm dS}_{\rm b}$.

Let us focus on the evolution of the tensor perturbations $h_{ij}$. For
convenience, we decompose the quantity $h_{ij}$ into the
three-dimensional spatial Fourier modes as
%
\begin{equation}
 h_{ij}(\tau,\mathbf{x},z) = \sum_{P}\int\!h^P_k(\tau,z)
          e^{i\mathbf{k}\cdot\mathbf{x}}\hat{e}^P_{ij}\,d^3k,
 \label{eq:Fourier}
\end{equation}
%
where $\hat{e}^P_{ij}$ denotes a transverse-traceless polarization
tensor. In terms of the Fourier modes, the evolution equation for the
perturbations is given by \cite{Koyama2004}
%
\begin{equation}
 \Box_{{\rm AdS}_5}h = \frac{\partial^2 h}{\partial \tau^2} 
                      - \frac{\partial^2 h}{\partial z^2} 
                      + \frac{3}{z}\frac{\partial h}{\partial z} 
                      + k^2h = 0.,
\label{eq:wave_equation}
\end{equation}
%
where $\Box_{{\rm AdS}_5}$ denotes the d'Alembertian operator in the
AdS$_5$ spacetime. Hereafter we omit subscripts of $h_k^P$ and simply
write $h$. The equation (\ref{eq:wave_equation}) must be solved with the
junction condition imposed on the brane. In the Poincar\'e coordinate,
the explicit form of the junction condition becomes \cite{Koyama2004}
%
\begin{equation}
 \left.\frac{\partial h}{\partial n}\right|_{\rm brane} =
  \left(\frac{\partial h}{\partial \tau} -
  \frac{\sqrt{1+H^2\ell^2}}{H\ell}\frac{\partial h}{\partial z}
  \right)_{z=z_{\rm b}(t)} = 0,
\label{eq:junction_condition}
\end{equation}
%
where $\partial/\partial n$ denotes the normal vector of the brane.
While there is generally a contribution from the tensor part of
anisotropic stress tensor on the brane $\pi_{ij}^{\rm T}$,
we neglect it for simplicity and set $\pi_{ij}^{\rm T} = 0$ hereafter.

It is important to note that the evolution equation (\ref{eq:wave_equation})
can be written in a separable form and by using this fact, one obtains
the general solutions (e.g. \cite{Gorbunov, Koyama2004})
%
\begin{equation}
h(\tau,z)=\int_0^{\infty} dm\,\,\left\{
\tilde{h}_1(m)\,z^2\,H_2^{(1)}(mz)\,e^{i\,\omega\,\tau} +
\tilde{h}_2(m)\,z^2\,H_2^{(2)}(mz)\,e^{-i\,\omega\,\tau} \right\},
\label{eq:general_solution} 
\end{equation}
%
where $\omega^2=m^2+k^2$. The functions $H_2^{(1)}$ and $H_2^{(2)}$ 
denote respectively the Hankel functions of first and second kind, and  
$\tilde{h}_{1}(m)$ and $\tilde{h}_{2}(m)$ represent arbitrary coefficients.
The above expression implies that the GWs propagating in the bulk are
described as a superposition of the zero mode ($m=0$) and the
KK-modes ($m>0$). Solving the wave equation (\ref{eq:wave_equation})
with the junction condition (\ref{eq:junction_condition}) is the
equivalent task to determining the coefficients $\tilde{h}_{1,2}(m)$ 
that satisfy the junction condition \cite{Koyama2004}. In the very
high-energy case, technical difficulties hinder efforts to calculate
these analytically because of the significant contribution from the
massive modes ($m\ell\gg 1$). For this reason, we solve numerically the
wave equation (\ref{eq:wave_equation}) with the junction condition
(\ref{eq:junction_condition}).

%
\subsection{Initial conditions}
\label{subsec:initial}
%

In order to correctly estimate the effects of the KK-modes, we must
specify the initial conditions for the perturbed quantity $h$ after the
inflation. In this paper, we specifically consider a brane inflation
model in which the exponential expansion takes place on the
brane. According to Ref. \cite{Langlois2000}, the GN coordinate system
$(t, \mathbf{x}, y)$ provides a useful spatial slicing in the
inflationary epoch. With this coordinate, the perturbed metric of the
AdS$_5$ spacetime becomes
%
\begin{equation}
 ds^2 = -N^2(y,t)dt^2+A^2(y,t)(\delta_{ij}+h_{ij})dx^idx^j+dy^2,
 \label{eq:metric_GN}
\end{equation}
%
where $A(t,y)=e^{Ht}N(y)$ and
$N(t,y)=N(y)=\cosh(y/\ell)- \left(1+\rho/\lambda\right)\sinh(y/\ell)$.
Note that the perturbations $h_{ij}$ satisfies the TT conditions
(\ref{eq:TTgauge}).
In this coordinate, our brane is located at a fixed point $y=0$.

During de Sitter inflation, the solution of the evolution equation
of GWs, $\Box_{{\rm AdS}_5}h=0$ [see Eq. (\ref{eq:wave_equation})], can be
obtained in a separable form, $h(t,y)\equiv u(y)\phi(t)$. Introducing
the separation constant $m$ which represents the mass of KK-modes, 
the mode functions
$u$ and $\phi$ satisfy the following equations \cite{Langlois2000}:
%
\begin{align}
&\frac{d^2\phi_m}{dt^2} + 3H\frac{d\phi_m}{dt} +
 \left(m^2+\frac{k^2}{a^2}\right)\phi_m = 0,\label{eq:separable_phi}\\
&\frac{d^2u_m}{dy^2} +
 4\frac{N'}{N}\frac{du_m}{dy}+\frac{m^2}{N^2}u_m=0,\label{eq:separable_u} 
\end{align}
%
where a prime denotes a derivative with respect to $y$.
Picking up the normalizable modes from the solutions of the equation
(\ref{eq:separable_u}), one notice that a large mass gap arises between 
the lightest KK-mode ($m=3H/2$) and the zero-mode ($m=0$)
\cite{Langlois2000}. From the point of view of the
quantum theory, the large mass gap highly suppresses the excitation of
KK-modes during inflation. Consequently, the zero-mode 
solution in the GN coordinates gives a dominant contribution to the
metric fluctuation \cite{Langlois2000}. Solving the equations
(\ref{eq:separable_phi}) and (\ref{eq:separable_u}) with $m=0$, the
zero-mode is described explicitly as 
%
\begin{equation}
 h(\eta, y) = C (-k \eta)^{3/2} H_{3/2}^{(1)}(-k \eta),
 \label{eq:DS_zero_mode}
\end{equation}
%
where $C$ is a normalization constant and $\eta$ is the conformal time
$\eta = -1/aH$. 

To see how the solution (\ref{eq:DS_zero_mode}) looks like in the
Poincar\'e coordinates, we rewrite the general solution
(\ref{eq:general_solution}) in the GN coordinates. In the de Sitter
case, the coordinate transformation between the GN coordinates and the
Poincar\'e coordinates is explicitly given as \cite{Gorbunov, Kobayashi}
%
\begin{equation}
 z = -\eta\sinh(y/\ell),\qquad \tau=\eta\cosh(y/\ell),
 \label{eq:GN_Poincare} 
\end{equation}
%
Substituting them into the general solution (\ref{eq:general_solution}),
we obtain 
%
\begin{eqnarray}
h(\tau,z) &=& 
\int_0^{\infty} dm \,\left\{\eta \sinh (y/\ell)\right\}^2 
\left\{ \tilde{h}_1(m) \,H_2^{(1)}\left(m \eta \sinh(y/\ell)\right)
\,\, e^{-i\,\omega \,\eta \cosh (y/\ell)} \right.
\nonumber\\
&& \left. ~~~~~~~~~~ 
+ \tilde{h}_2(m) \,H_2^{(2)}\left(m \eta \sinh (y/\ell)\right) 
\,\, e^{i\,\omega \,\eta \cosh (y/\ell)} \right\}. 
\label{eq:DeSitter} 
\end{eqnarray}
%
Comparing (\ref{eq:DS_zero_mode}) with (\ref{eq:DeSitter}), 
we see that the
zero-mode solution given in the inflationary epoch cannot be simply
expressed in terms of the zero-mode  solution in the Poincar\'e
coordinates, which indicates that a mixture of KK-modes is required to 
express the zero-mode solution in the inflationary epoch. Nevertheless, 
in the long-wavelength limit $k\to0$, both the zero-mode solutions 
become constant over the time and the bulk space, and they coincide with 
each other. Since we are specifically concerned with the evolution of 
long-wavelength GWs after inflation, the constant mode, i.e., 
$h=\text{const.}$ and $dh/d\tau=0$, seems a natural and a physically 
plausible initial condition for our numerical calculation in the 
Poincar\'e coordinate. Strictly speaking, however, this is valid only in
the long-wavelength limit $k\to0$. This point will be discussed in
details in Sec \ref{subsec:bulk}.

\section{Numerical simulation}
\label{sec:numerical}

%
\subsection{Numerical scheme}
\label{subsec:scheme}
%

On the basis of the formalism presented in the previous section, we now
discuss the numerical treatment used to solve the wave equation
(\ref{eq:wave_equation}) with the junction condition
(\ref{eq:junction_condition}). First of all, the computational domain
should be finite. We introduce an artificial cutoff
(regulator) boundary in the bulk at $z=z_{\rm reg}$ (shown in
Fig. \ref{fig:Poincare}) and impose the Neumann condition at the
regulator boundary,  i.e.,
%
\begin{equation}
\left(\frac{\partial h}{\partial z}\right)_{z=z_{\rm reg}}=0.\label{eq:neumann}
\end{equation}
%
In this paper, numerical calculations were carried
out by employing the pseudo-spectral method \cite{Canuto}. The amplitude
of GWs $h(\tau, z)$ is decomposed in terms of Tchebychev polynomials,
defined as
%
\begin{equation}
T_n(\xi)\equiv \cos(n\cos^{-1}(\xi))\;\;{\rm for}\;\; -1\leq \xi \leq 1,
\end{equation}
%
which yields a polynomial function of $\xi$ of order $n$. For example,
%
\begin{equation}
T_0(\xi) = 1,\quad T_1(\xi) = \xi,\quad T_2(\xi)=2\xi^2-1,\cdots
\end{equation}
%
Here the variable $\xi$ is related to the Poincar\'e coordinate z. To
implement the pseudo-spectral method, instead of using the Poincar\'e
coordinates $(\tau, z)$ directly, we use the new coordinates $(t,\xi)$:
%
\begin{equation}
\tau=T(t),\qquad
z=\frac{1}{2}\left[\,\left\{z_{\rm reg}-z_{\rm b}(t)\right\}\xi + 
\left\{z_{\rm reg}+z_{\rm b}(t)\right\} \,\, \right]
\label{eq:Tchebychev_coordinate}
\end{equation}
%
so that the locations of both the physical and the regulator branes 
are kept fixed, and the spatial coordinate $z$ is projected to the
compact domain $-1\leq \xi \leq 1$. Adopting this coordinate system, the
amplitude $h(t,\xi)$ is first transformed into the Tchebychev space
through the relation,
%
\begin{equation}
h(t,\xi)= \sum^N_{n=0} \widetilde{h}_n(t)T_n(\xi),
\label{eq:Tchebychev_transform}
\end{equation}
%
where we set $N=2048$ or $4096$. We then discretize the $\xi$-axis to
the $N+1$ points (collocation points)  using the inhomogeneous grid
$\xi_n=\cos(n\pi/N)$ called Gauss-Lobatto collocation
points. With this grid, fast Fourier transformation can be applied to
perform the transformation between the amplitude $h(t,\xi)$ and the
coefficients $\widetilde{h}_n(t)$. Then the wave equation
(\ref{eq:wave_equation}) rewritten in the new coordinates are decomposed
into a set of ordinary differential equations (ODEs) for
$\widetilde{h}_n(t)$. For the temporal evolution of
$\widetilde{h}_n(t)$, we use Adams-Bashforth-Moulton method with the
predictor-corrector scheme. Further technical details of the numerical
scheme is summarized in Appendix \ref{app:pseudospectral}.

%
\subsection{Setup and parameters}
\label{subsec:setup}
%

We are especially concerned with the late-time evolution of GWs after
the inflation. For this purpose, we focus on the evolution equation
(\ref{eq:wave_equation}) in the RD phase. In order to quantify
high-energy effects, we define a useful parameter
$\epsilon_*$ which represents the normalized energy density at the
horizon re-entry time $t_*$ of the GWs concerned, namely,
$\epsilon_*=\epsilon(t_*)$. From the Friedmann equation
(\ref{eq:Friedmann}) and the definition of the scale factor
(\ref{eq:scalefactor_energy}), the comoving wave number
of GWs is rewritten in terms of the parameter $\epsilon_*$ as
%
\begin{equation}
 k=a_*H_*=\left(\frac{\epsilon_{\rm crit}}{\epsilon_*}\right)^{1/3(1+w)}
   \sqrt{\epsilon_*^2+2\epsilon_*}.
\end{equation}
%
For higher frequency GWs, $\epsilon_*$ becomes larger. One thus
expects that the high-frequency GW modes tend to be significantly
affected by the high-energy effects.

Notice that the location of the regulator brane is another important
parameter. Here, the location of the boundary is set to
$z_{\rm reg} = 25$--$200\ell$, which is far enough away from the
physical brane to avoid artificial suppression of light KK-modes. 
Further, we must stop the numerical calculations before the influence of
the boundary condition at $z=z_{\rm reg}$ reaches the physical brane
$z_{\rm b}$. The arrival time of the influences of the regulator brane
can be estimated by drawing a null line from the initial position of the
regulator boundary $(\tau_0, z_{\rm reg})$ toward the physical
brane. With these treatments, we have checked that the amplitude of GWs
on the brane is fairly insensitive to the location of regulator
boundaries. Thus, all the results presented in Sec. \ref{sec:results1}
and Sec. \ref{sec:results2} are free from the effect of regulator boundary.

In the situation considered here, the initial time $t_{\rm init}$ is
also an important parameter, which turns out to have an important effect
on the GWs in the bulk \cite{Hiramatsu2}. We parameterize the initial
time as
%
\begin{equation}
 s_{\rm init} \equiv \frac{a(t_{\rm init})H(t_{\rm init})}{k},
 \label{eq:s_init}
\end{equation}
%
which represents the wavelength of GWs normalized by the Hubble horizon
scale at the initial time $t_{\rm init}$. In order to get a reliable
estimate, we set $s_{\rm init} \gg 1$ and run the simulations until
$\epsilon(t) \ll 1$, when the high-energy effects on the GWs become
negligible.  

Finally, we adopt the constant mode $h(t_{\rm init},\xi)=1$ as an
initial condition according to the discussion in Sec. \ref{subsec:initial}. 
Although it is plausible, the validity of the constancy of the
superhorizon modes must be checked. This point will be carefully
discussed in Sec. \ref{subsec:bulk}. 

\section{Code check and qualitative behavior of GWs}
\label{sec:results1}

%
\subsection{Code check}
\label{subsec:codetest}
%

In order to check our numerical code, we first consider the simplest case
in which the location of the brane is given by $z=z_{\rm b}=$constant.
This is the so-called Minkowski brane embedded in the AdS$_5$ bulk.
The solution of the evolution equation (\ref{eq:wave_equation}) is given
as \cite{RS2,Sasaki1999,Gorbunov}
%
\begin{equation}
 h_{\rm exact}(\tau,z) = z^2Z_2(mz)E(\omega\tau),
 \label{eq:Minkowski_solution}
\end{equation}
%
where $\omega \equiv \sqrt{k^2+m^2}$. The function $Z_2$ and $E$
respectively denote linear combinations of the Bessel functions of order
2 and the sinusoidal functions. Imposing the junction condition
(\ref{eq:junction_condition}) at $z_{\rm b}=\ell$, we obtain 
%
\begin{equation}
 h_{\rm exact}(\tau,z) = \left(\frac{z}{\ell}\right)^2\left\{Y_1(m)J_2(mz) 
                         - J_1(m)Y_2(mz)\right\}\cos(\omega\tau).
 \label{eq:Minkowski_exact_solution}
\end{equation}
%
According to the analytic solution (\ref{eq:Minkowski_exact_solution}),
we set $h=h_{\rm exact}(0,z)$ and 
$\dot{h}=\partial h_{\rm exact}/\partial\tau|_{\tau=0}=0$
for the initial condition of numerical simulation and compare the
numerical results with the analytic solution.
 
\begin{figure}[ht]
 \centering
 {\includegraphics[width=9cm]{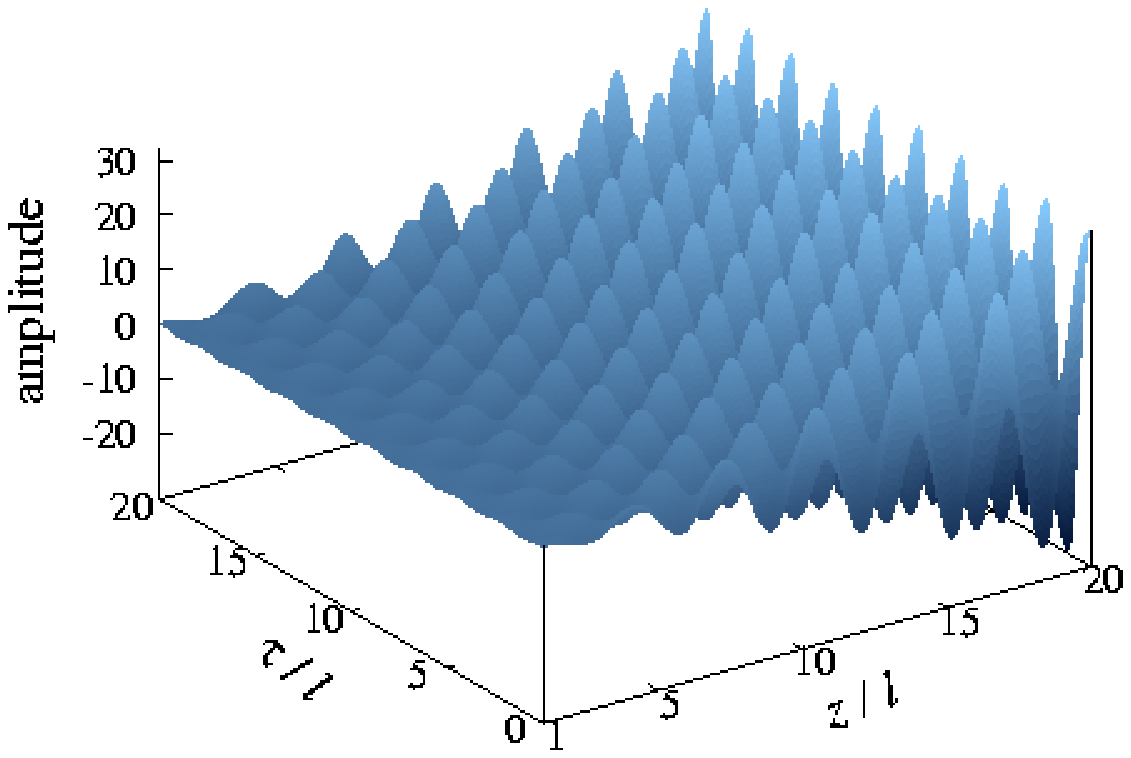}
 \includegraphics[width=8cm]{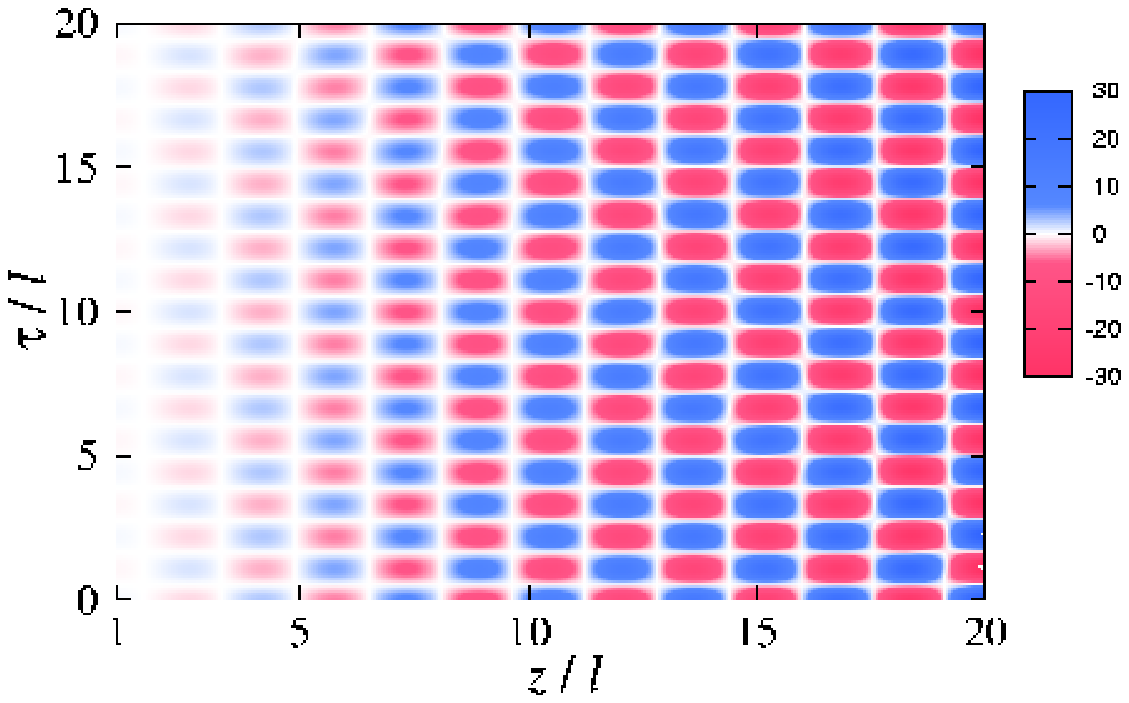}}
 \caption{The behavior of a test wave with $m\ell=k\ell=2$ in the bulk.
 The Minkowski brane is located at $z=\ell$. The right panel depicts the
 projection of the three-dimensional waves of the left panel. 
 \label{fig:codecheck1}} 
\end{figure}

\begin{figure}[ht]
 \centering
 \includegraphics[width=16cm]{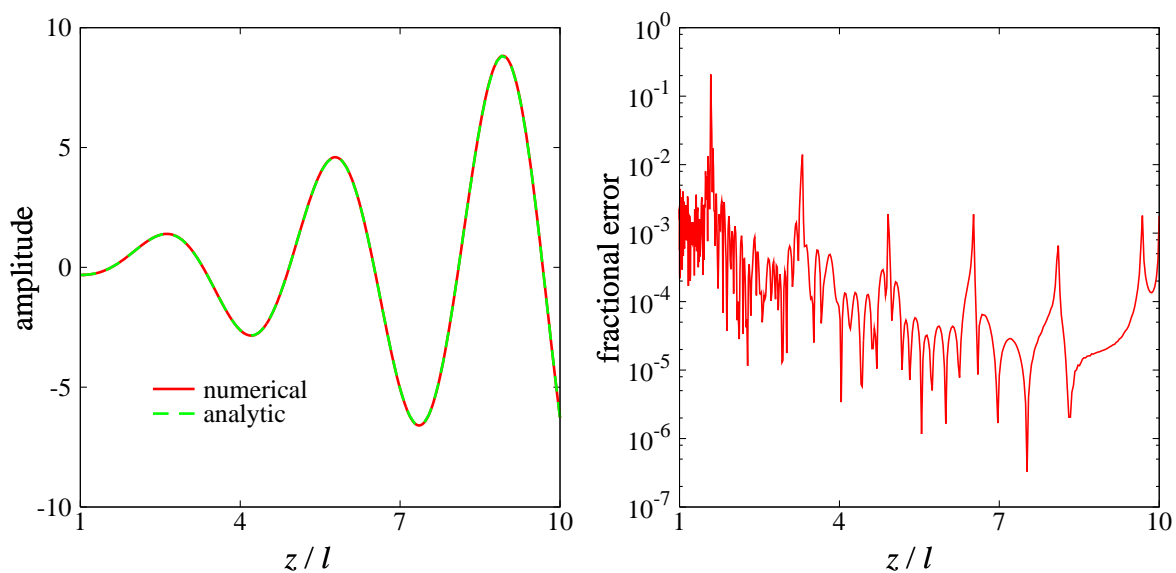}
 \caption{The left panel illustrates that the numerical solution at
 $\tau=10$ is consistent with the analytical solution
 (\ref{eq:Minkowski_exact_solution}) in the region $z_{\rm b}\leq z\leq 10$.
 The outer region $10 \leq z \leq z_{\rm reg}$ in the bulk is
 contaminated by the boundary condition on the regulator brane at
 $z=z_{\rm reg}$. The right panel shows the numerical errors
 $|(h_{\rm num}-h_{\rm exact})/h_{\rm exact}|$ estimated at that
 time, which is suppressed by $10^{-3}$ near the physical
 brane. The spike-shapes reflects the cancellation of
 significant digits because of $h_{\rm exact}(10,z)\approx 0$.
 \label{fig:codecheck2}}
\end{figure}

Fig. \ref{fig:codecheck1} shows the behavior of GWs in the AdS$_5$
bulk. The right panel of the figure is the projection of the left
panel. In this simulation, we chose parameters as $z_{\rm reg}=20\ell$,
$m\ell=k\ell=2$ and $N=1024$. The left panel of
Fig. \ref{fig:codecheck2} shows the snapshot of the waveform at
$\tau=\tau_1=10\ell$, which illustrates that the numerical result
accurately recovers the exact solution
(\ref{eq:Minkowski_exact_solution}) in the interval between
$z_{\rm b}\leq z \leq z_{\rm reg}-\tau_1$. Outside this region, the
numerical simulation is contaminated by the boundary condition of the
regulator brane. In the right panel of Fig. \ref{fig:codecheck2},
the fractional error of the amplitude 
$|(h_{\rm num}-h_{\rm exact})/h_{\rm exact}|$ evaluated at the time
$\tau=\tau_1$ is plotted as a function of bulk coordinate. We found that
the error is suppressed to the order of $10^{-3}$ near the physical
brane. Note that there appear several sharp spikes, whose locations
roughly match those of the zero-point $h_{\rm exact}(\tau,z)=0$. Thus,
the cancellation of significant digits occurs. These numerical errors
can be reduced when the number of collocation points $N$ is increased.


%
\subsection{Behavior of GWs in the bulk and the validity check of the
  initial condition}
\label{subsec:bulk}
%

Having checked the reliability of our numerical scheme, we now focus on
the cosmological evolution of GWs. As we mentioned in Sec
\ref{subsec:initial}, we must first clarify the validity and the
sensitivity of the initial condition, namely, the constancy of the
superhorizon modes. It should be stressed that, only in the
long-wavelength limit $k\to0$ during the inflationary phase, the
constant mode coincides with the zero-mode solution
[see Eq.(\ref{eq:DeSitter})]. Therefore, the constancy of GW amplitudes
after inflation cannot be guaranteed even on superhorizon
scales. Depending on the choice of the parameters $s_{\rm init}$ and
$\epsilon_*$, the mode $h=$ const. may not be a good approximation to
the initial condition for numerical simulations in the RD epoch.

\begin{figure}[ht]
 \centering
 {\includegraphics[width=9cm]{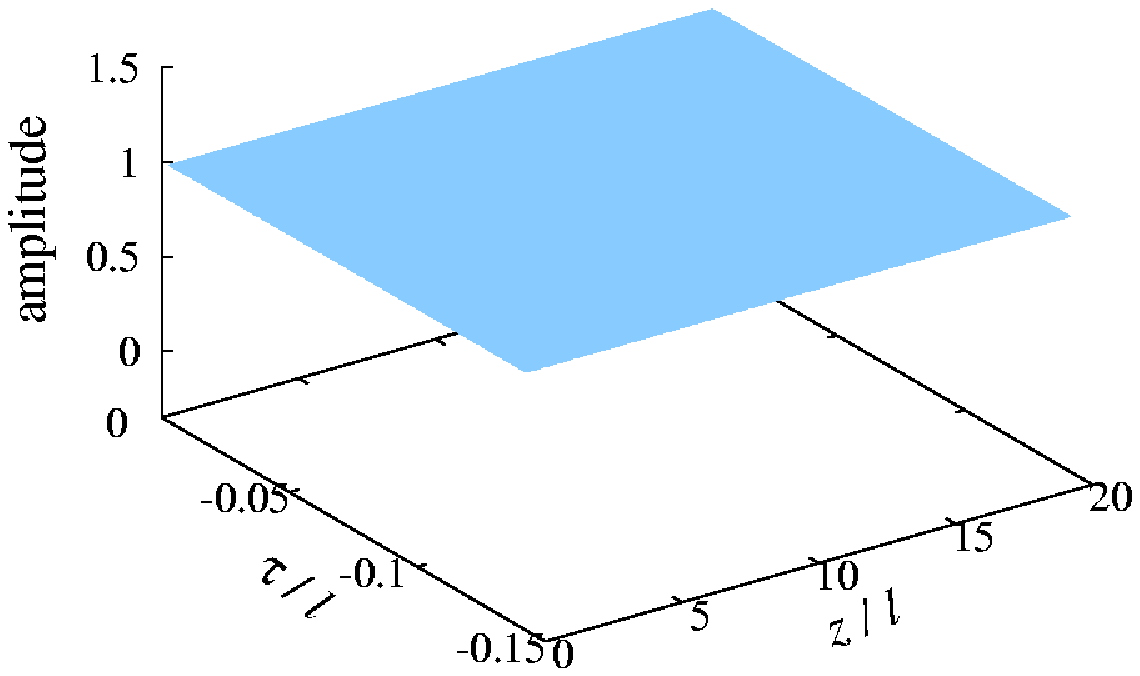}
 \includegraphics[width=8cm]{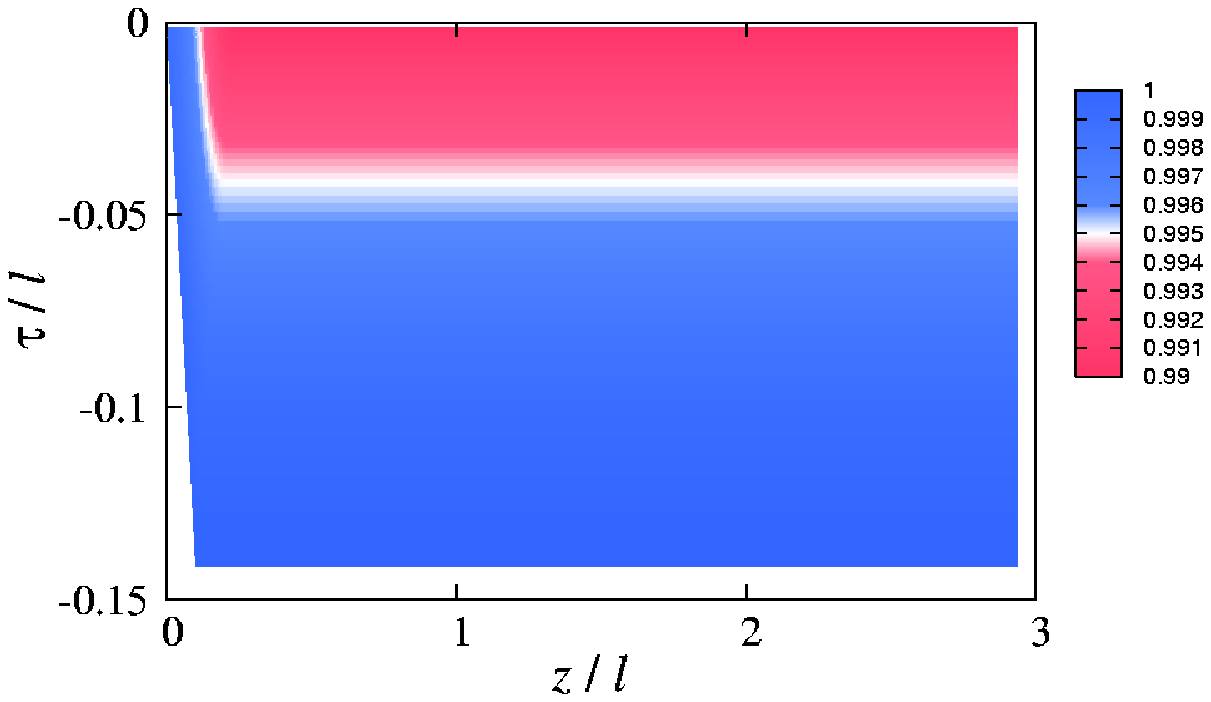}}
 \caption{The evolution of a GW in the bulk in the case of a de Sitter
 brane. We set the Hubble parameter to $H\ell=\sqrt{3}$
 with $(s_{\rm init},z_{\rm reg})=(10,20)$. 
 The right panel depicts the projection of the three-dimensional waves
 of the left panel, zooming in the image in $0\leq z/\ell\leq 3$.
 The empty corner in the surface represents the motion of the brane
 [see Eq.(\ref{eq:DS_trajectory})].
 \label{fig:5D_behavior_DS}}
\end{figure}

Fig. \ref{fig:5D_behavior_DS} shows the time evolution of GWs in the
Poincar\'e coordinate system in the de Sitter case. In this simulation,
we set $s_{\rm init}=100$ and $H\ell=\sqrt{3}$. The universe on the
brane experiences
accelerated cosmological expansion and the wavelength of GWs becomes longer
than the Hubble horizon. Fig. \ref{fig:5D_behavior_DS}
indicates that the GWs on initially superhorizon-scale remains constant
not only on the brane but also in the bulk. The right panel of this
figure, which depicts the projection of the left panel, shows that a
very slight change of the amplitude is observed (a fraction of the
original amplitude of $\leq1\%$) and the amplitude finally converged to
a fixed value. In this sense, the constant mode $h=const$. is suitable
for the initial condition of the superhorizon-scale GWs in the RD phase
if we impose the initial condition just after the inflation.

\begin{figure}[ht]
 \centering
 {\includegraphics[width=9cm]{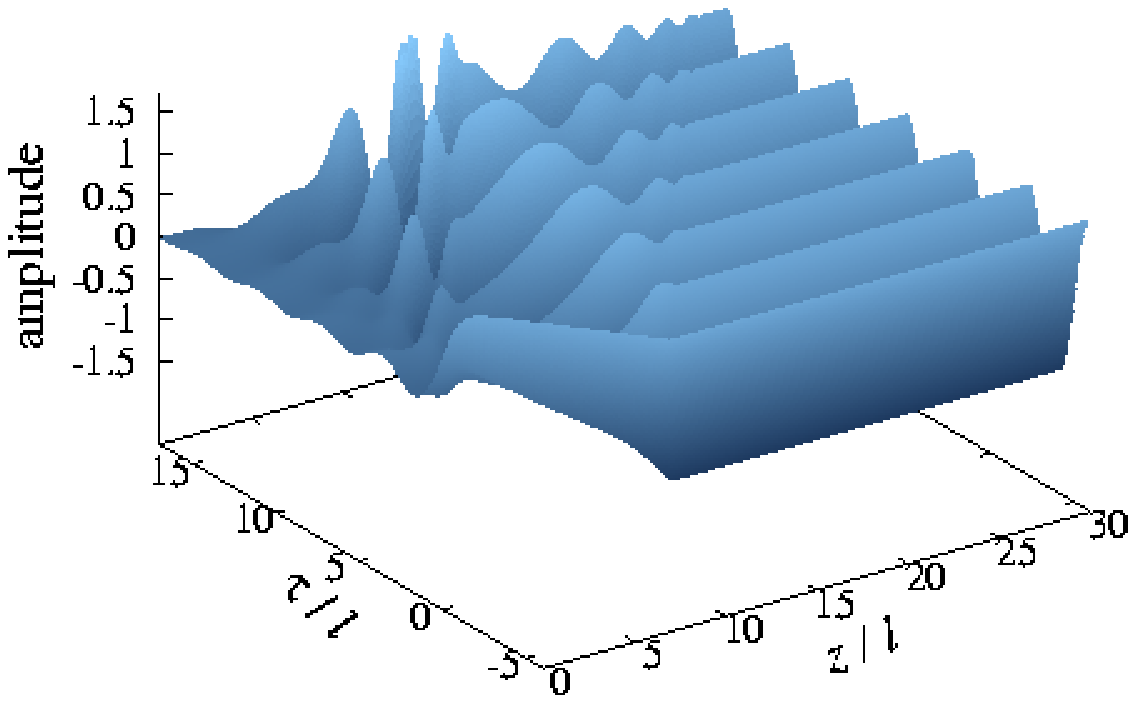}
 \includegraphics[width=8cm]{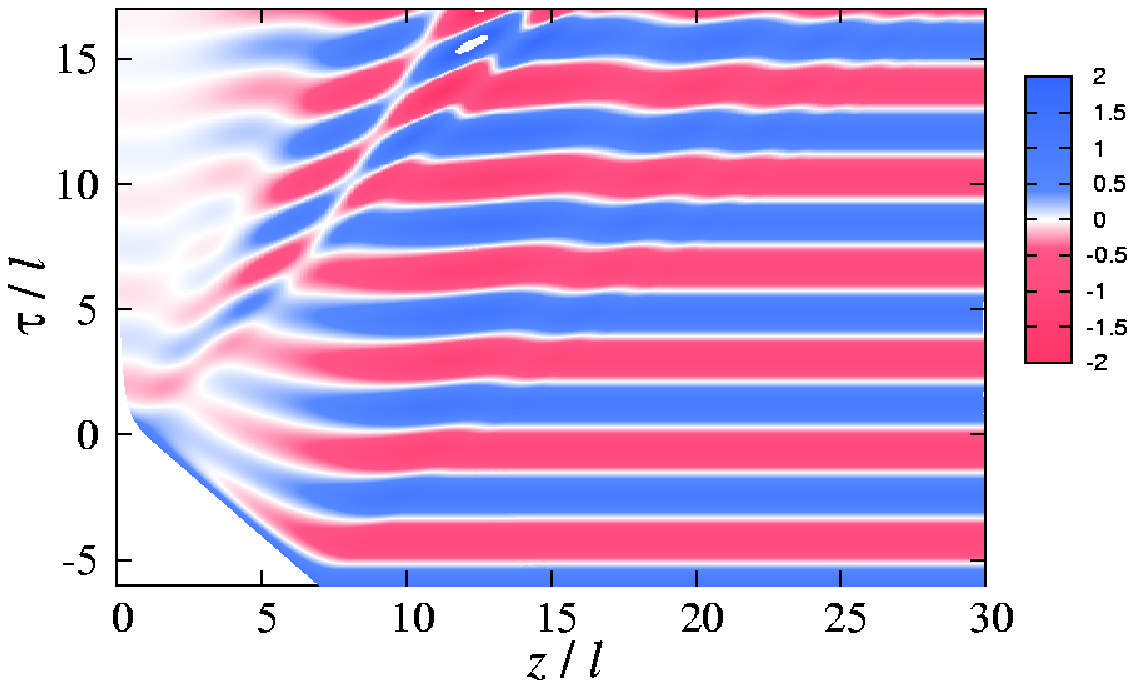}}
 \caption{The evolution of a GW in the bulk in the case of a
 Friedmann brane. We set the comoving wave number to 
 $k=\sqrt{3}/\ell$ or $\epsilon_*=1.0$ with 
 $(s_{\rm init},z_{\rm reg})=(200,80)$. 
 The right panel depicts the projection of the three-dimensional waves
 of the left panel. The empty corner in the surface represents the
 motion of the brane [see Eq.(\ref{eq:brane_trajectory})].
 \label{fig:5D_behavior_FRW}} 
\end{figure}

Adopting this initial condition, we then performed simulations in the
radiation-dominated FRW case $(w=1/3)$ with same parameters,
$\epsilon_*=1.0$ and $s_{\rm init}=100$. This result is shown in 
Fig \ref{fig:5D_behavior_FRW}. We found that the constancy of
the GW amplitude no longer holds in the bulk even before $\tau\approx 0$,
where the wavelength of GW on brane just re-enters the Hubble horizon.
In particular, GWs emanating from the physical brane are observed, which
propagate into the bulk spacetime almost along a null line. This
indicates that the excitation of KK-modes occurs near the brane
even if the wavelength of GWs is still outside the Hubble horizon. 

It is noteworthy that the different behaviors of GWs in the AdS$_5$ bulk
may be caused by the difference in the motion of the brane
[see Eqs.(\ref{eq:brane_trajectory}) and (\ref{eq:DS_trajectory})].
In the moving mirror problem in an electromagnetic field, the
acceleration or deceleration of the mirror yields the creation of
photons due to vacuum polarization (e.g., Sec. 4.4 of
Ref. \cite{BirrelDavies}). A similar phenomenon may occur in the
AdS$_5$ bulk, that is, the KK-modes (massive gravitons) are excited by
the deceleration of the brane which is depicted by an arc with a
non-zero curvature $d^2z_{\rm b}/d\tau^2<0$.

Figs. \ref{fig:5D_behavior_DS} and \ref{fig:5D_behavior_FRW} reveal
that the constant mode $h(t_{\rm init},\xi)=\text{const.}$ can
be used as the initial condition if we set this just after the end of
inflation, but, the constancy of the long-wavelength mode would not
be guaranteed in the RD epoch even on superhorizon scales.
This implies that the choice of the initial time $t_{\rm init}$ (or
$s_{\rm init}$) defined in (\ref{eq:s_init}) is crucial when setting the
initial condition at the RD epoch.  

\begin{figure}
 \centering
 \includegraphics[width=12cm]{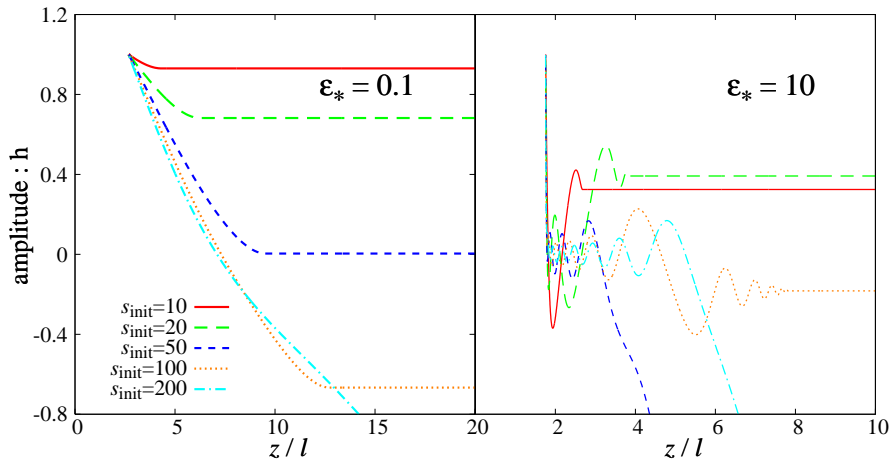}
\caption{Snapshots of the GW amplitudes in the bulk for various choices
 of initial time. The snapshots were taken when the
 wavelength of GWs becomes five times longer than the Hubble horizon,
i.e., $aH/k=5$. \label{fig:sensitivity_to_s0_bulk}}
\end{figure}

\begin{figure}
 \centering
 \includegraphics[width=12cm]{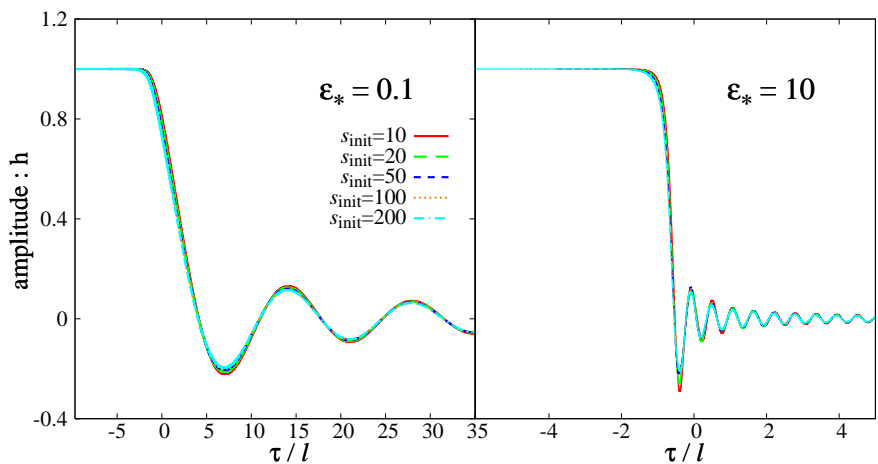}
\caption{Evolved results of GWs projected on the brane 
starting with the various initial times.
 \label{fig:sensitivity_to_s0_brane}}
\end{figure}

In Figs. \ref{fig:sensitivity_to_s0_bulk} and
\ref{fig:sensitivity_to_s0_brane}, the dependence of the evolution of
GWs on the initial time is shown by varying the parameter
$s_{\rm init}$ in low-energy ($\epsilon_*=0.1$) and high-energy
($\epsilon_*=10$) cases. Fig. \ref{fig:sensitivity_to_s0_bulk} plots the
snapshots of the amplitude $h(\tau,z)$ in the bulk when the wavelength
of GWs becomes five times larger than the Hubble horizon, i.e.,
$aH/k=5$. Clearly, in the bulk, the amplitude of GWs is very sensitive
to the choice of the parameter $s_{\rm init}$, or equivalently, the
initial time $t_{\rm init}$. The resultant wave-form away from the
physical brane does not show any convergence even in the low-energy case
$(\epsilon_*=0.1)$. This behavior may be caused by the fact that the
constant mode with the comoving wave number $k$ in the RD epoch
immediately starts to oscillate as $h(\tau,z)\propto e^{ik\tau}$, which
is the massless ($m\to0$) limit of Eq. (\ref{eq:general_solution}).

On the other hand, in Fig. \ref{fig:sensitivity_to_s0_brane},
the GW amplitudes tend to converge on the brane if we set the initial
time $t_{\rm init}$ early enough. This convergence property might be due
to the presence of the junction condition (\ref{eq:junction_condition}).
Therefore, as far as we choose $s_{\rm init} \gtrsim 50$ for our
interest of the energy scale $0.01\lesssim \epsilon_*\lesssim 100$, we
do not need to care about the initial time, when we estimate the IGWB
spectra on the brane. In Appendix \ref{app:initial_time}, quantitative
aspects of the convergence properties of the amplitude are discussed.
Moreover, Seahra addressed these points in an analytic way
in Ref.\cite{Sanjeev2006}.


\section{IGWB spectra}
\label{sec:results2}

%
\subsection{Comparison with reference waves}
\label{subsec:reference}
%

Keeping the results in Sec. \ref{sec:results1} in mind, let us now
quantitatively estimate the high-energy effects of the GWs and evaluate
the energy spectra of the IGWB on the brane. To quantify these, it might
be helpful to discriminate the influence of KK-mode excitation in the
bulk from the non-standard cosmological expansion caused by the
$\rho^2$-term in the Friedmann equation (\ref{eq:Friedmann}). For this
purpose, we introduce the reference wave $h_{\rm ref}$, which is a
solution of the four-dimensional evolution equation of the amplitude
obtained by replacing the scale factor and the Hubble parameter derived from
the standard Friedmann equation with those from the modified Friedmann
equation (\ref{eq:Friedmann}). The resultant equation is given by
%
\begin{equation}
  \ddot{h}_{\rm ref}+3H\dot{h}_{\rm ref}+\left(\frac{k}{a}\right)^2
h_{\rm ref}=0,
  \label{eq:reference}
\end{equation}
%
which is just the Klein-Goldon equation for a scalar field in the FRW
background (e.g. \cite{Allen1988,Maggiore}) and is same as
(\ref{eq:separable_phi}) for $m=0$. Comparing the numerical 
simulations with the solution of the wave equation (\ref{eq:reference}),
the effect of the KK-mode excitation can be quantified separately.

\begin{figure}
 \centering
 \includegraphics[width=16cm]{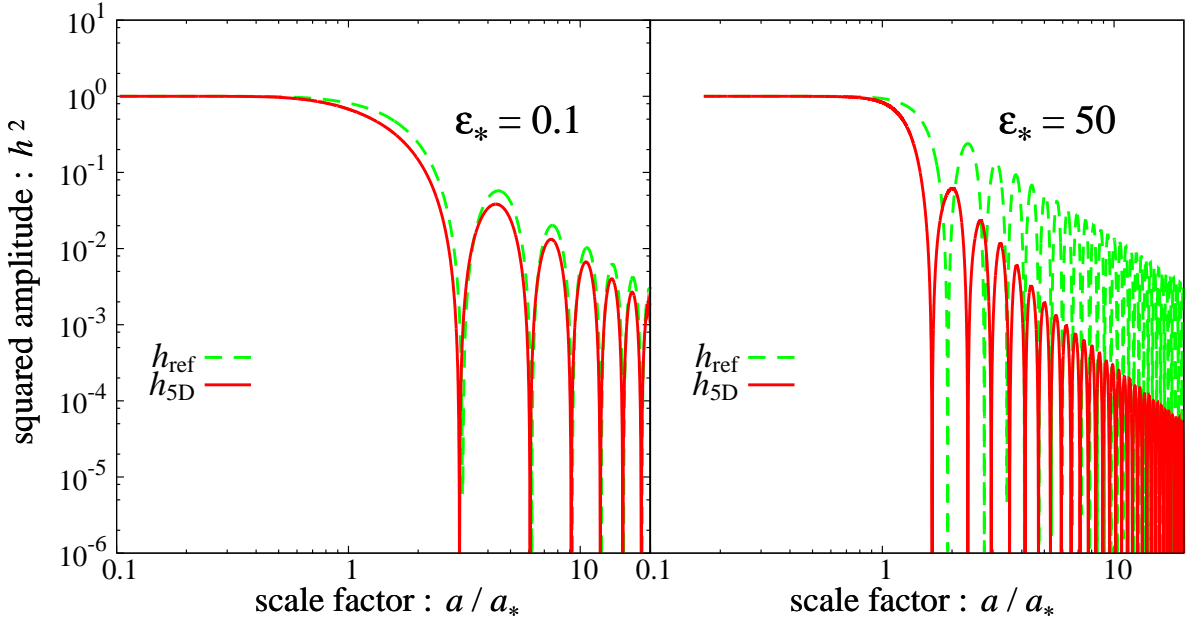}
 \caption{Squared amplitude of GWs on the brane in low-energy ({\it
 left}) and the high-energy ({\it right}) regimes. In both panels, solid 
 lines represent the numerical solutions of wave equation
 (\ref{eq:wave_equation}). The dashed lines are the amplitudes of
 reference wave $h_{\rm ref}$ 
 obtained from equation (\ref{eq:reference}). 
\label{fig:brane_behavior}}
\end{figure}

Fig. \ref{fig:brane_behavior} shows the squared amplitude of the GWs,
$h_{\rm 5D}^2$ and $h_{\rm ref}^2$ as functions of the scale factor
$a$. The left panel shows the low-energy case ($\epsilon_*=0.1$), 
while the right panel depicts the result in the high-energy regime 
($\epsilon_*=50$). As we increase the energy scale at the horizon
re-entry time, the GW amplitude $h_{\rm 5D}$ becomes considerably reduced
compared to the reference wave, $h_{\rm ref}$. Since the late-time
evolution of GWs simply scales as $h\propto 1/a$ in both $h_{\rm 5D}$
and $h_{\rm ref}$, the suppression of the amplitude $h_{\rm 5D}$ is
caused by the excitation of KK-modes around the horizon re-entry time. 
Notice that the normalized energy density at the horizon re-entry time
$\epsilon_*$ is related to the observed proper frequency
$2\pi\,f=k/a_{0}=(a_*/a_{0})H_*$ as 
%
\begin{equation}
\frac{f}{f_{\text{crit}}}
   =\left(\frac{a_*}{a_{\rm crit}}\right)\ell H_* 
   =\left(\frac{\epsilon_{\rm crit}}{\epsilon_*}\right)^{1/3(1+w)}
                             \sqrt{\epsilon_*^2+2\epsilon_*},
  \label{eq:relation_with_eps}
\end{equation}
%
where the critical frequency $f_{\text{crit}}$ is defined in
(\ref{eq:critical_frequency}) as 
$2\pi f_{\text{crit}}=(a_{\rm crit}/a_{0})\ell^{-1}$, and $w=1/3$
in this case. From this relation, one expects that the KK-mode excitation
is essential in the high-energy regime and the deviation from the
standard four-dimensional prediction for the spectrum of IGWB becomes
more prominent above the critical frequency, $f>f_{\rm crit}$. 

%
\subsection{IGWB spectrum in the five-dimensional cosmology}
\label{subsec:spectrum}
%

We are in position to estimate the influence of KK-mode excitation on
the shape of the spectrum. To do so, we ran simulations for the
parameters listed in Table \ref{tab:parameters} ($w=1/3$ case) and
estimated the ratio of amplitudes $|h_{\rm 5D}/h_{\rm ref}|$ for a
different set of parameters. Note that in the simulations with
$\epsilon_* \lesssim 1$, the location of regulator brane $z_{\rm reg}$
should be set far away from the physical brane. This is because the
long-term evolution is needed to follow the oscillatory behavior. 

\begin{figure}
 \centering
 \includegraphics[width=8cm]{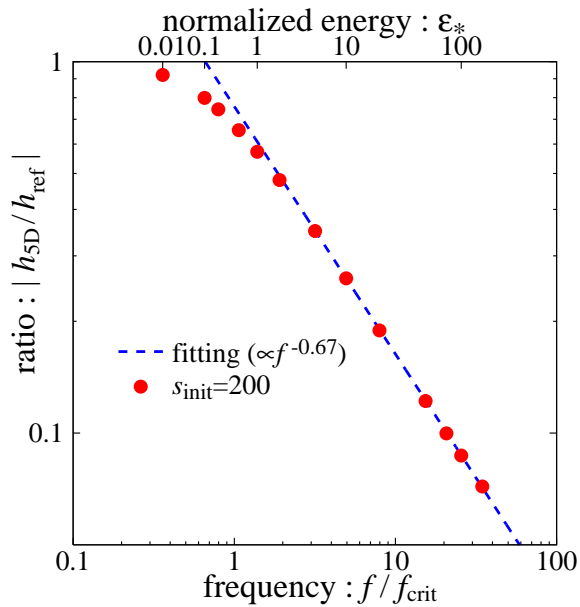}
 \caption{Frequency dependence of the ratio of amplitudes 
 $|h_{\rm 5D}/h_{\rm ref}|$ between the numerical simulation of wave 
 equation (\ref{eq:wave_equation}) and the reference wave
 (\ref{eq:reference}). The vertical solid line represents the critical
 frequency. The dashed line indicates the fitting result
 (\ref{eq:fitting}), where fitting was examined using the data with
 $s_{\rm init}=200$ at the asymptotic region $\epsilon_* \geq 5$. 
 \label{fig:amplitude_ratio}}
\end{figure}

We show the frequency dependence of the ratio in
Fig. \ref{fig:amplitude_ratio}. The ratio is evaluated at the low-energy
regime long after the horizon re-entry time and is plotted as a function
of the frequency $f/f_{\rm crit}$. Clearly, the ratio 
$|h_{\rm 5D}/h_{\rm ref}|$ monotonically decreases with the frequency
and the suppression of amplitude $h_{\rm 5D}$ becomes significant above
the critical frequency $f_{\rm crit}$. Using the data points in the
asymptotic region $\epsilon_*\geq 5$, we try to fit the ratio of
amplitudes with $s_{\rm init}=200$ to a power-law function. Employing
the least-squares method, the result becomes
%
\begin{equation}
  \left|\frac{h_{\rm 5D}}{h_{\rm ref}}\right| = 
\alpha\left(\frac{f}{f_{\text{crit}}}\right)^{-\beta}
  \label{eq:fitting}
\end{equation}
%
with $\alpha=0.76\pm0.01$ and $\beta=0.67\pm0.01$ (dashed line in
Fig. \ref{fig:amplitude_ratio}). In Appendix \ref{app:initial_time},
we calculate the ratios for various $s_{\rm init}$ for each combination
$(\epsilon_*,z_{\rm reg})$ to check the robustness of this result.

\begin{table}
 \centering{
  \begin{tabular}{c||c|c|c}
  \hline
  $w$ & $(\epsilon_*,z_{\rm reg})$ & $s_{\rm init}$ & $N$ \\ \hline\hline
 0 & $(0.1,200\ell)\sim(100,50\ell)$&$50,100,200,400$ & 2048 or 4096 \\
 $1/3$& $(0.01,200\ell)\sim(100,25\ell)$&$50,100,200,400,800$ & 2048 or 4096 \\
 1 & $(0.1,500\ell)\sim(50,50\ell)$&$200,400,800,3200,12800$ & 2048 or 4096 \\
  \hline
 \end{tabular}
 }
 \caption{Numerical parameters used for the simulations to estimate the
 frequency dependence of the ratio of amplitudes $|h_{\rm 5D}/h_{\rm ref}|$.
 }
 \label{tab:parameters}
\end{table}

\begin{figure}
 \centering
 \includegraphics[width=8cm]{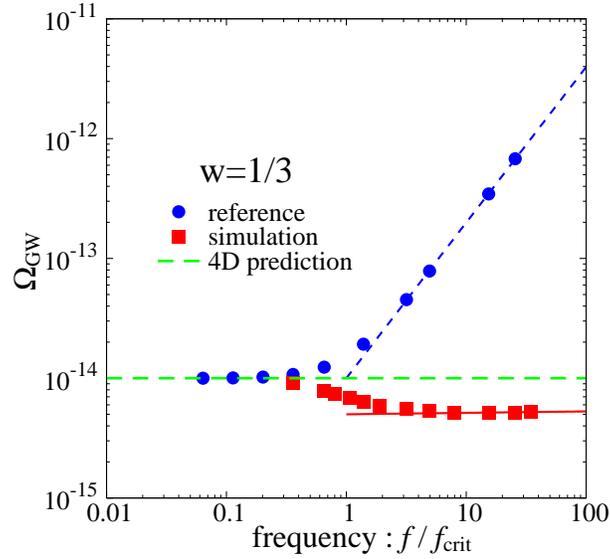}
 \caption{The energy spectrum of the IGWB around the critical
 frequency. The filled circles represent the spectrum caused by
 the non-standard cosmological expansion of the
 universe. Taking account of the KK-mode excitations, the spectrum
 becomes the one plotted as filled squares. Particularly, in the
 asymptotic region depicted in the solid line, the frequency dependence
 becomes almost same as the one predicted in the four-dimensional theory
 (long-dashed line).\label{fig:spectrum_RD}}
\end{figure}

The power-law fit (\ref{eq:fitting}) can be immediately translated to
the energy spectrum of IGWB, $\Omega_{\rm GW}$.
The spectrum taking account of the KK-mode excitations is calculated as
%
\begin{equation}
  \Omega_{\rm GW} =\left|\frac{h_{\rm 5D}}{h_{\rm
		    ref}}\right|^2\Omega_{\rm ref},
  \label{eq:Omega_GW_ref}
\end{equation}
%
where we used the fact $\Omega_{\rm GW}\propto h^2f^2$.
As discussed in Sec. \ref{subsubsec:high-energy}, if we neglect the
effect of the KK-mode excitation, the spectrum becomes
$\Omega_{\rm ref}\propto f^{4/3}$
[See Eq.\ref{eq:O_propto_f_specific_5D}]. Then, combining 
it with the result (\ref{eq:fitting}), the IGWB
spectrum becomes nearly flat above the critical frequency :
%
\begin{equation}
  \Omega_{\rm GW} \propto f^{0},
  \label{eq:fitting_Omega}
\end{equation}
%
which is shown in filled squares in Fig. \ref{fig:spectrum_RD}.
In this figure, the spectrum calculated from the results of the
reference waves $\Omega_{\rm ref}$ is also shown in filled circles.
Note that the normalization factor of the spectrum is determined so as
to be $\Omega_{\rm GW} = 10^{-14}$ according to the constraint from the
CMB observation. The short-dashed line and the solid line represent each 
asymptotic behavior in the high-frequency region. The spectrum taking
account of the two high-energy effects seems almost indistinguishable
from the standard four-dimensional prediction shown in long-dashed line
in the figure. In other words, while the effect due to the non-standard
cosmological expansion lifts up the spectrum, the KK-mode effect reduces
the GW amplitude, which results in the same spectrum as the one
predicted in the four-dimensional theory. Additionally notice that the
amplitude taking account of the two effects near $f\approx f_{\rm crit}$
is slightly decreasing, which agrees with the results in our previous
study for $\epsilon_*\leq0.3$ using the GN coordinates \cite{Hiramatsu1}.

At this point, however, it is unclear whether the results obtained here
is generic or accidental for certain range of the model parameters. To
clarify the cosmological dependence of the KK-mode excitations in more
quantitative way, we next study the cases with different EOS
parameter $w$.

%
\subsection{Dependence on equation of state}
\label{subsec:EOS}
%

To quantify the EOS dependence, we ran
simulations for the MD case $w=0$ and the somewhat stiff matter case
$w=1$, which might be realized by introducing the kinetically driven
scalar field (e.g. the quintessential inflation \cite{Sahni2002, Sami2004}).
Varying the EOS parameter changes the acceleration of the brane. 
One naively expects that the different motion of the brane may suppress
or enhance the KK-mode excitation.

\begin{figure}
 \centering
 \includegraphics[width=8cm]{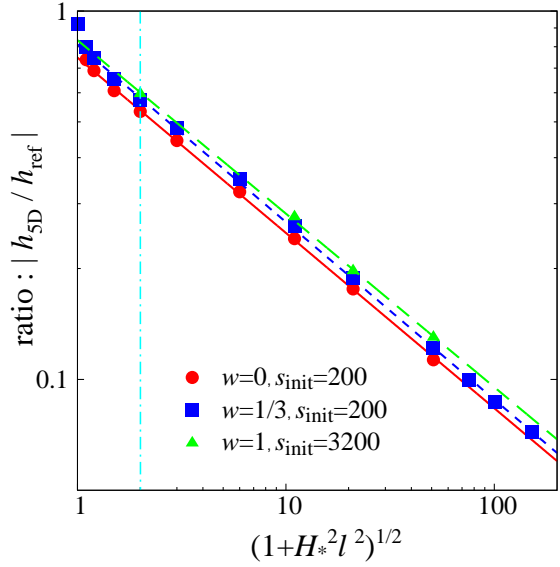}
 \caption{EOS dependence on the amplitude ratio. The filled circles,
 squares and triangles show the cases with $w=0, 1/3$ and $1$,
 respectively. The solid line, short-dashed line and long-dashed line
 show each fitting result above the critical frequency corresponding to
 $(1+H_{\rm crit}^2\ell^2)^{1/2}=\sqrt{2}$ depicted in the vertical
 dot-dashed line. \label{fig:EOS}}
\end{figure}

With the same procedure as in the previous subsection, we calculated the
ratio of amplitudes $|h_{\rm 5D}/h_{\rm ref}|$ for various $\epsilon_*$
in the case of $w=0$ and $w=1$. The results are summarized in
Fig. \ref{fig:EOS}, where the horizontal axis represents
$\sqrt{1+H_*^2\ell^2}$. Fitting the power-law function to all cases
shown in the figure, we found that the ratios universally scale as 
%
\begin{equation}
  \left|\frac{h_{\rm 5D}}{h_{\rm ref}}\right|=
   \widetilde{\alpha}(1+H_*^2\ell^2)^{-0.24}
   \approx \widetilde{\alpha}(1+H_*^2\ell^2)^{-1/4},
  \label{eq:universal_relation1}
\end{equation}
%
where $\widetilde{\alpha}=0.75, 0.81$ and $0.83$ for $w= 0, 1/3$ and 1,
respectively,  which indicates that the quantity $\widetilde{\alpha}$
may be related to $w$ as $\widetilde{\alpha}=-0.155 w^2+0.241w+0.748$ 
by simply fitting the quadratic function. An important point to
emphasize is that this scaling property does not depend on the parameter
$w$ or the acceleration of the brane. Combining the scaling relation
with (\ref{eq:Omega_GW_ref}), the IGWB spectra can be estimated as
%
\begin{equation}
  \Omega_{\rm GW} = \widetilde{\alpha}^2(1+H_*^2\ell^2)^{-1/2}\Omega_{\rm ref}.
  \label{eq:universal_relation2}
\end{equation}
%
Especially, in the high-frequency region $f\gg f_{\rm crit}$, the
prefactor of the right-hand side behaves as
%
\begin{equation}
  (1+H_*^2\ell^2)^{-1/2} \approx H_*\ell 
                \propto f^{-\frac{3(w+1)}{3w+2}}
  \label{eq:universal_relation3}
\end{equation}
%
from equations (\ref{eq:H_propto_t}),(\ref{eq:f_propto_t}) and
(\ref{eq:scalefactor_index}). Combining the result (\ref{eq:Omega_w}), 
the energy spectrum of the IGWB behaves as
%
\begin{equation}
 \Omega_{\rm GW} \propto
  f^{\frac{3w-1}{3w+2}} \;\; {\rm for}\; f \gg f_{\rm crit}.
 \label{eq:IGWB_spectra}
\end{equation}
%

Owing to these calculations (\ref{eq:Omega_w}) and (\ref{eq:IGWB_spectra}),
for $w=0$ (MD), we obtain 
%
\begin{equation}
 \Omega_{\rm GW} \propto
 \begin{cases}
  f^{-1/2} & {\rm for}\;\; f \gg f_{\rm crit}, \\
  f^{-2}   & {\rm for}\;\; f \ll f_{\rm crit},
 \end{cases} \label{eq:asymptotic_MD}
\end{equation}
%
and for $w=1$ case,
%
\begin{equation}
 \Omega_{\rm GW} \propto
 \begin{cases}
  f^{2/5} & {\rm for}\;\; f \gg f_{\rm crit}, \\
  f^{1}   & {\rm for}\;\; f \ll f_{\rm crit}.
 \end{cases} \label{eq:asymptotic_KD}
\end{equation}
%
These results imply that the spectrum generally changes from
the four-dimensional prediction. Indeed, transforming the numerical
results (\ref{eq:universal_relation1}) to the energy spectra in the
cases with $w=0$ and $w=1$, the frequency dependence of the spectra
including the two high-energy effects (solid lines) clearly differ from
each four-dimensional prediction (long-dashed lines). 

Taking these results into account, one may conclude that the
cancellation of the high-energy effects in the RD epoch is accidental
and the KK-mode excitation dominates over the non-standard cosmological
expansion when $w>1/3$.

\begin{figure}
 \centering
 {\includegraphics[width=8cm]{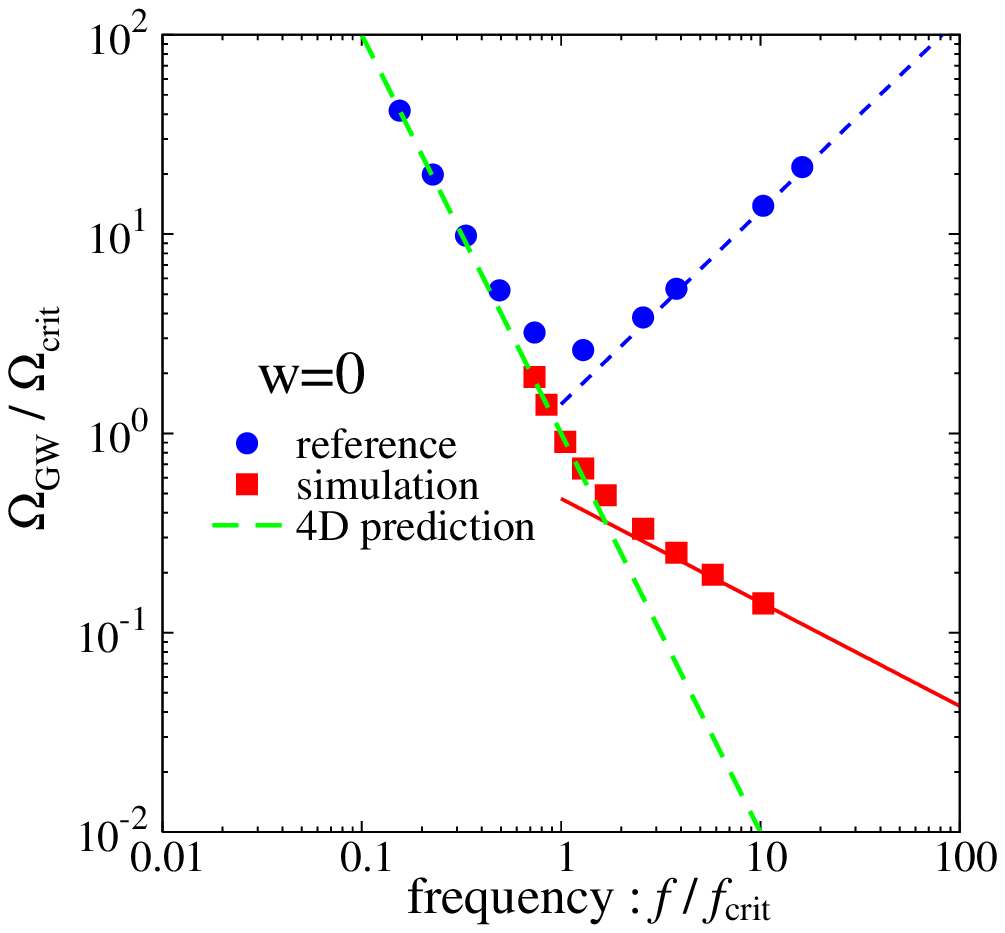}
  \includegraphics[width=8cm]{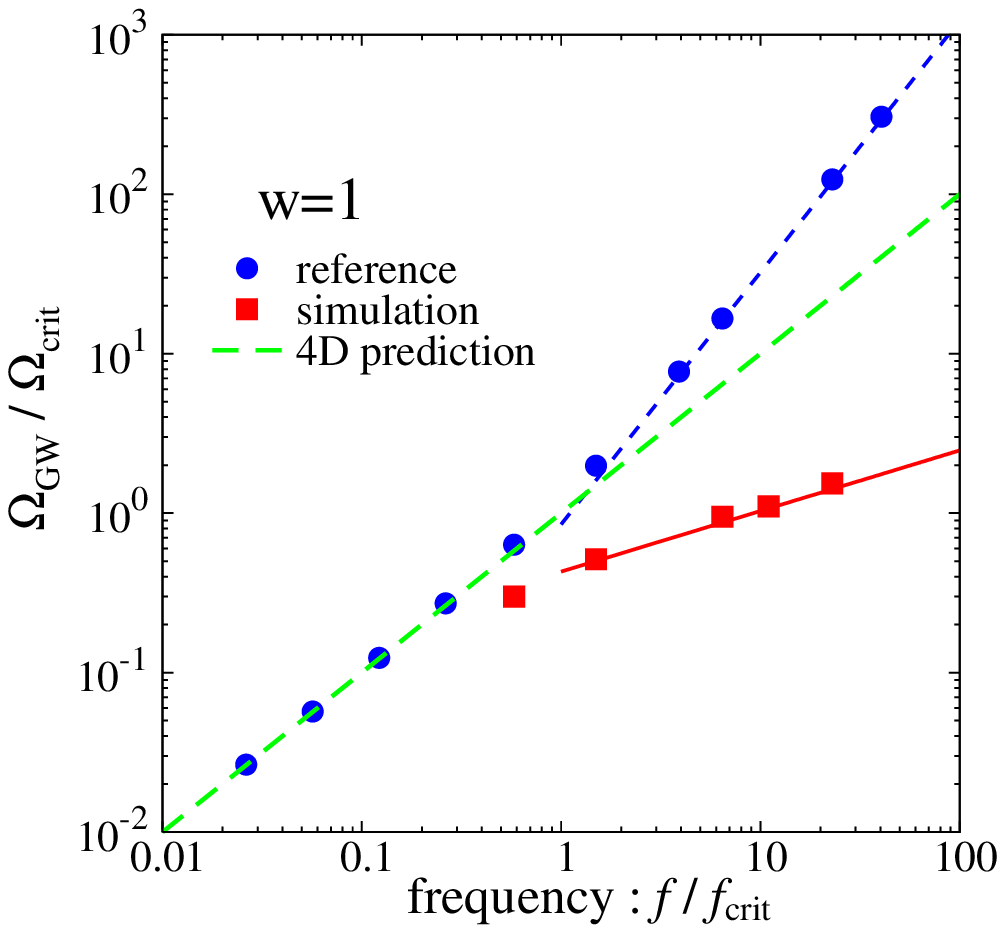}}
 \caption{The energy spectrum of the IGWB in the background EOS with
 $w=0$(left), and with $w=1$(right). The amplitude is normalized by the
 value of the four-dimensional prediction at the critical frequency.
  \label{fig:spectrum_EOS}}
\end{figure}

\section{Conclusion}
\label{sec:conclusion}

We have investigated the power spectrum of the IGWB in the
five-dimensional cosmology based on the Randall-Sundrum model.
In the braneworld scenario, the two high-energy effects affect the shape of
the spectrum above the critical frequency $f_{\rm crit}$ defined in 
(\ref{eq:critical_frequency}). One is the non-standard cosmological
expansion on the brane caused by the high-energy correction of the
Friedmann equation. The analytical estimate taking account of this
effect reveals that the effect makes the spectrum steeply blue
[see Eq. (\ref{eq:O_propto_f_specific_5D})]. By contrast, another
important effect is the excitations of KK-modes which escapes from our
brane into the five-dimensional bulk, leading to the suppression of the
spectrum. In order to quantify these two effects, we solved the wave
equation of each Fourier mode of GWs numerically for various EOS
parameters $w$.

The systematic survey of numerical simulations with various parameter
sets reveals that there may exist the universal scaling law 
for the KK-mode excitation in the high-energy regime
[Eq.(\ref{eq:universal_relation1})]:
%
\begin{equation*}
  \left|\frac{h_{\rm 5D}}{h_{\rm ref}}\right|
   \propto (1+H_*^2\ell^2)^{-1/4}.
\end{equation*}
%
Using the universal scaling law, we constructed the power spectrum of
the IGWB in the cases with $w=0$ (MD universe), $w=1/3$ (RD universe)
and $w=1$ (stiff matter dominant universe). From the results
(\ref{eq:Omega_w_4D}) and (\ref{eq:IGWB_spectra}), the frequency
dependence of the spectrum in the high-frequency region $f>f_{\rm crit}$
becomes
%
\begin{equation}
  \Omega_{\rm GW} \propto 
  \begin{cases}
    f^{-2}\;({\rm 4D})\;,\;f^{-1/2}\;({\rm 5D}) &{\rm for}\;\; w=0,\\
    f^{0},\;\hspace{3em} \;f^{0}                &{\rm for}\;\; w=1/3,\\
    f^{1},\;\hspace{3em} \;f^{2/5}              &{\rm for}\;\; w=1.
  \end{cases}
\end{equation}
%
Particularly, in the RD case, the accidental
cancellation of the two high-energy effects occurs, which yields the
same spectrum as one predicted in the four-dimensional (4D) theory. 
This scaling law might be understood in a context of moving mirror
problems. The discussion about the analytic derivation of the scaling
law is work in progress.

Finally, we briefly comment on the other numerical works using the
different schemes. Recently, Seahra solved numerically the
wave equation in the null coordinate system based on the Poincar\'e
coordinates using a sophisticated numerical scheme
\cite{Sanjeev2006}. He observed the agreement between his numerical
results and our results for the same choice of the EOS parameters with
the same initial conditions as ours.¡¡In addition, it is observed that,
even if a constant initial condition on the initial null hypersurface is
chosen, the amplitude of GWs on the brane is almost identical with our
results. Moreover, the numerical calculation based on the quantum theory
has been performed by Kobayashi and Tanaka \cite{Kobayashi2005B}. They
reported the same spectrum in the RD case as ours, even if KK modes are
taken into account in the initial de Sitter phase. On the other hand,
Ichiki and Nakamura have obtained a tilted 
spectrum $\Omega_{\rm GW}\propto f^{-0.46}$ \cite{Ichiki}. 
While their early results have included errors associated with numerical
accuracy, the new calculation using the revised code did not converge to
the flat spectrum either. Currently, we do not know the reason why the
result by Ichiki and Nakamura is different from ours. In order to
understand these numerical results well, the analytical study of the
scaling relation (\ref{eq:universal_relation1}) is essential, which is
definitely our next task. 

\begin{acknowledgments}  
I would like to thank Atsushi Taruya, Sanjeev Seahra and Kazuya
Koyama for carefully reading this manuscript and giving me useful
comments. TH is supported by JSPS (Japan Society for the Promotion of
Science).
\end{acknowledgments}    

\appendix  

\section{The pseudo-spectral method}
\label{app:pseudospectral}

In this appendix, we briefly describe the implimentation of the
pseudo-spectral method in our numerical scheme.

From the coordinate transformation (\ref{eq:Tchebychev_coordinate}),
the wave equation (\ref{eq:wave_equation}) is expressed as 
%
\begin{align}
 \frac{\partial^2h}{\partial t^2}
   +K_{t\xi}\frac{\partial^2h}{\partial t\partial \xi}
   +K_{\xi\xi}\frac{\partial^2h}{\partial \xi^2}
   +K_{t}\frac{\partial h}{\partial t}
   +K_{\xi}\frac{\partial h}{\partial \xi}
   +Kh=0,
\end{align}
%
where the coefficients $K_{t\xi}, K_{\xi\xi}, K_{t}, K_{\xi}$ and $K$
are functions of $t$ and $\xi$. We use the predictor-corrector
method for the temporal evolution. To implement this, we introduce an
auxiliary variable $\chi(t,\xi)$ satisfying the equation  
%
\begin{align}
 \frac{\partial h}{\partial t} = \chi -
             K_{t\xi}\frac{\partial h}{\partial \xi}
             \equiv F(\chi,h';t,\xi_n),
 \label{eq:def_F}
\end{align}
%
where the prime denotes the derivative with respect to $\xi$
[see Eq.(\ref{eq:Tchebychev_coordinate})].
With this definition, the time evolution of $\chi$ satisfying to
(\ref{eq:wave_equation}) is formally written as
%
\begin{align}
 \frac{\partial \chi}{\partial t} = G(\chi,h,h',h'';t,\xi_n).
 \label{eq:def_G}
\end{align}
%
Notice that the function $G$ does not contain the derivative
$\partial\chi/\partial \xi$. Empirically, the presence of this
derivative causes numerical instability.
The functions $F$ and $G$ are evaluated at each collocation point $\xi_n$.
Then, transforming to the Tchebychev space by
Eq. (\ref{eq:Tchebychev_transform}), we obtain a set of ordinary
differential equations :
%
\begin{equation}
 \frac{d\widetilde{h}_n}{dt}    = \widetilde{F}_n(t),\;\;
 \frac{d\widetilde{\chi}_n}{dt} = \widetilde{G}_n(t).\label{eq:reduced_wave}
\end{equation}
%
With the reduced equations (\ref{eq:reduced_wave}), 
the predictor-corrector method based on the Adams-Bashforth-Moulton
scheme can be used to obtain the time evolution of $\widetilde{h}_n(t)$
and $\widetilde{\chi}_n(t)$.

At each time step, while the boundary conditions
(\ref{eq:junction_condition}) and (\ref{eq:neumann}) are used to
evaluate $\widetilde{h}_{N-1}$ and $\widetilde{h}_{N}$, we put
additional conditions on $\widetilde{\chi}_{N-1}$ and
$\widetilde{\chi}_{N}$ as 
%
\begin{equation}
 \widetilde{\chi}_{N-1} = \widetilde{\chi}_{N} = 0.\label{eq:boundary_G}
\end{equation}
%
Owing to the definition of the function $F$ containing no derivatives of
$\chi$ [see Eq.(\ref{eq:def_F})], this empirically based treatment of
boundary conditions suppresses the numerical error caused by finite
truncation of the Tchebychev transformation (\ref{eq:Tchebychev_transform}).

In summary, we firstly evaluate the functions $F$ and $G$ in the
physical space at the time $t$. Then, transforming them into the
Tchebychev space by (\ref{eq:Tchebychev_transform}), 
we obtain $\widetilde{h}_n$ and $\widetilde{\chi}_n$ at the next time
step $t+\Delta t$ by solving the reduced equations (\ref{eq:reduced_wave})
for $0\leq n\leq N-2$, and imposing the boundary conditions and the
additional conditions (\ref{eq:boundary_G}) into $\widetilde{h}_{n}$ and
$\widetilde{\chi}_{n}$ for $n=N-1,N$. The spectral coefficients of the
derivatives $h'$ and $h''$ can be computed in decreasing order by the
recurrence relations \cite{Canuto}
%
\begin{align}
 c_n\widetilde{h}_n^{(1)} = \widetilde{h}_{n+2}^{(1)} +
 2(n+1)\widetilde{h}_{n+1}, \qquad \widetilde{h}_{n\geq N}^{(1)} = 0,
\end{align}
%
and then
%
\begin{align}
 c_n\widetilde{h}_n^{(2)} = \widetilde{h}_{n+2}^{(2)} +
 2(n+1)\widetilde{h}_{n+1}^{(1)}, \qquad \widetilde{h}_{n\geq N}^{(2)} = 0, 
\end{align} 
%
where $c_n$ is defined as
%
\begin{align}
 c_n = 
 \begin{cases}
  2 & {\rm for}\;\;n=0,N \\
  1 & {\rm for}\;\;1\leq n\leq N-1
 \end{cases}.
\end{align} 
%
Finally, we obtain $h(t+\Delta t,\xi)$ and its derivatives
$h'(t+\Delta t,\xi)$ and $h''(t+\Delta t,\xi)$ by the inverse Tchebychev
transformation.

\section{Initial time dependence}
\label{app:initial_time}

As seen in \ref{subsec:bulk}, the amplitudes on the brane becomes
insensitive to the choice of the initial time as long as $s_{\rm init}$
is large enough. In this appendix, we show that the ratios of amplitudes
discussed in \ref{subsec:spectrum} tend to converge to a fixed value
in the low-energy and high-energy cases. This validates the estimation of
the power spectrum of the IGWB using the results obtained from the
simulations with $s_{\rm init}=200$. 

\begin{figure}
 \centering
 {\includegraphics[width=8cm]{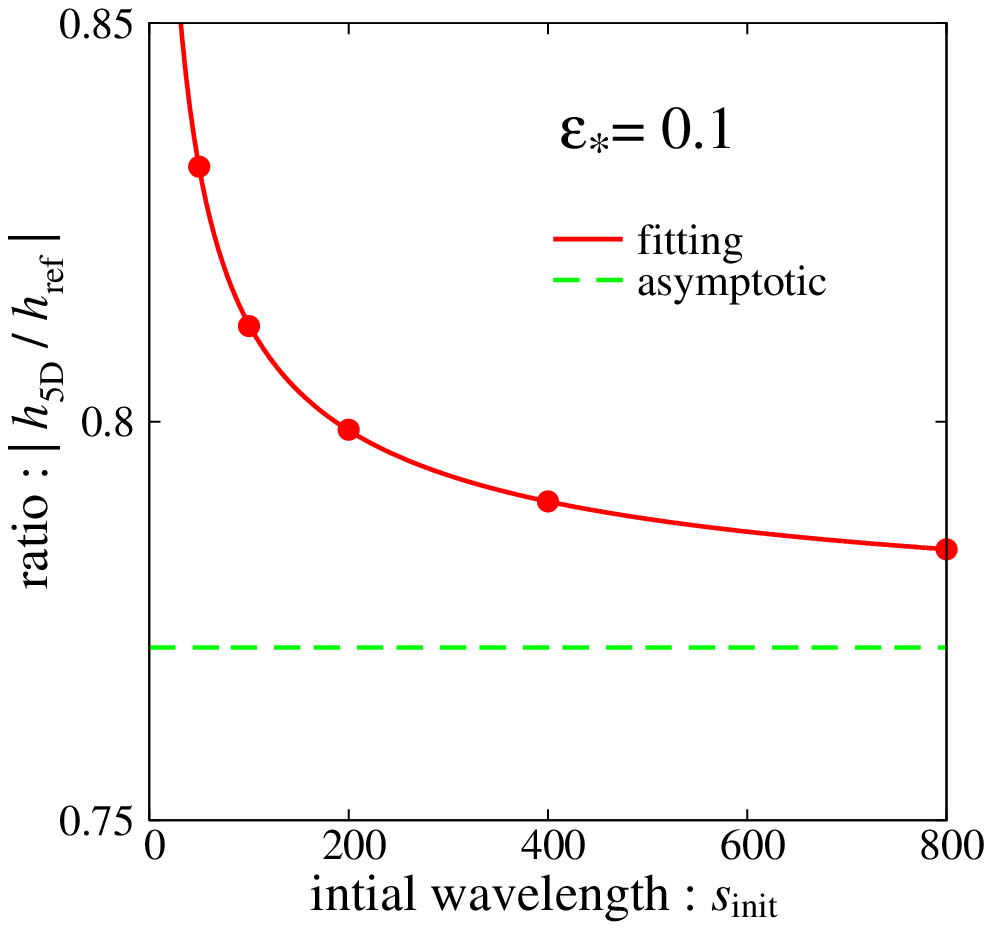}
  \includegraphics[width=8cm]{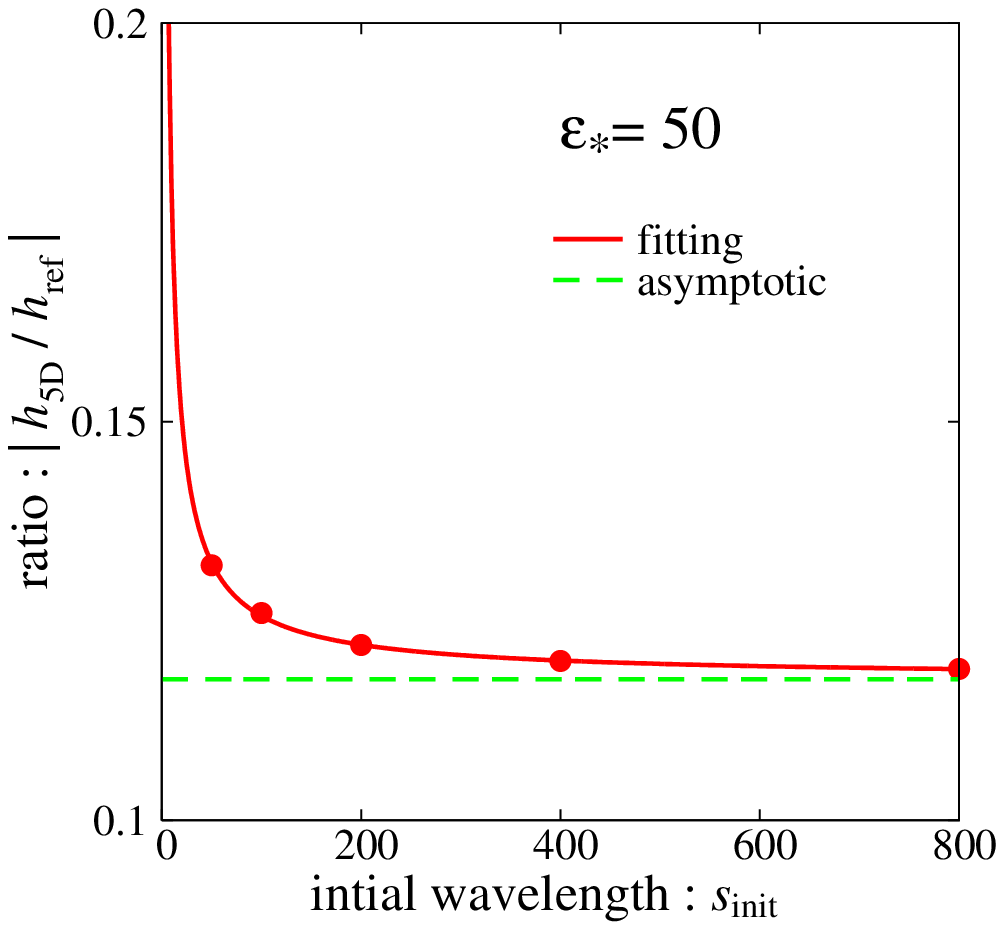}}
\caption{Convergence of the ratio of amplitudes 
 $h_{\rm 5D}/h_{\rm ref}$. Solid lines represent the fitting curve
 calculated by employing the nonlinear least-square method. Long-dashed
 lines denote the asymptotic value of the ratio.
\label{fig:sensitivity_to_s0}}
\end{figure}

Fig. \ref{fig:sensitivity_to_s0} shows the dependence of
the ratios $h_{\rm 5D}/h_{\rm ref}$ on the parameter $s_{\rm init}$. 
The left and right panels show the high-energy ($\epsilon_*=50$) and the
low-energy ($\epsilon_*=0.1$) cases, respectively. In both cases, the
ratios clearly converge to certain asymptotic values as increasing
$s_{\rm init}$. Using the nonlinear least-square method, we tried to fit
these values to the function
%
\begin{equation}
  \left|\frac{h_{\rm 5D}}{h_{\rm ref}}\right| = A s_{\rm init}^{B}+C,
\end{equation}
%
where $A,B,C$ are fitting parameters depending on $\epsilon_*$.
Then we obtained
%
\begin{equation}
  (A,B,C) = \begin{cases}
  (0.566,\;-0.573,\;0.772) &{\rm for}\;\;\;\epsilon_*=0.1,\\
  (0.456,\;-0.880,\;0.118) &{\rm for}\;\;\;\epsilon_*=50,
 \end{cases}
\end{equation}
%
which are shown in solid curves in Fig. (\ref{fig:sensitivity_to_s0}).
Note that $C$ represents the asymptotic values shown in long-dashed
lines in each panel.

Picking up the values at $s_{\rm init}=200$ in both cases, one can see
that the deviations from the asymptotic values $C$ keep to be less than
a few percent. From this fact, we use the ratios with
$s_{\rm init}=200$ to construct the power spectra of the IGWB without
deriving the asymptotic values for each $\epsilon_*$.

\bibliographystyle{apsrev}

\end{document}